\documentclass[epj]{svjour}
\usepackage{graphics}
\usepackage{amsmath}
\usepackage{amssymb}
\usepackage{amsfonts}
\usepackage{cuted}
\usepackage{lipsum} 
\usepackage{bibspacing}
\usepackage[utf8]{inputenc}
\usepackage{here}
\usepackage{graphicx}  
\usepackage{dcolumn}   
\usepackage{bm}        
\usepackage{amssymb}   
\usepackage{amsmath}   
\usepackage{subcaption}
\usepackage{color}
\usepackage{cite}

\usepackage[utf8]{inputenc}
\usepackage[T1]{fontenc}

\usepackage{url}
\usepackage[colorlinks, citecolor=blue,anchorcolor=blue,menucolor=blue, 
linkcolor=blue,filecolor=blue,runcolor=blue,urlcolor=blue,frenchlinks=true]{hyperref}

\definecolor{orange}{rgb}{0.93, 0.57, 0.13}
\definecolor{purple}{rgb}{0.75, 0.0, 1.0}

\newcommand{\red}[1]{\textcolor{red}{#1}}
\newcommand{\orange}[1]{\textcolor{orange}{#1}}
\newcommand{\yellow}[1]{\textcolor{yellow}{#1}}
\newcommand{\green}[1]{\textcolor{green}{#1}}
\newcommand{\purple}[1]{\textcolor{purple}{#1}}
\newcommand{\cyan}[1]{\textcolor{cyan}{#1}}
\newcommand{\blue}[1]{\textcolor{blue}{#1}}

\newcommand{\largeNc}{\red{L}\orange{a}\yellow{r}\green{g}\cyan{e} $\blue{\text{N}}_{\purple{\text{c}}}$ }

\begin{document}
\title{\centerline{The \largeNc limit of QCD on the lattice}} 

\titlerunning{\,\,The large $N_c$ limit of QCD on the lattice}
\authorrunning{P. Hern\'andez, F. Romero-L\'opez}

\author{\centerline{ \bf Pilar Hern\'andez\thanks{m.pilar.hernandez@uv.es} \and Fernando Romero-L\'opez\thanks{fernando.romero@uv.es}}}
\institute{\centerline{ Instituto de F\'isica Corpuscular, Universitat de Val\`encia and CSIC }}

\date{\hspace*{6.4cm} {\today}\vspace{1cm}}
%
\abstract{We review recent progress in the study of the large $N_c$ limit of gauge theories from lattice simulations. The focus is not only the planar limit but also the size of 
${\mathcal O}(N_c^{-1})$ corrections for values of $N_c\gtrsim 3$.  
Some concrete examples of the topics we include are tests of large-$N_c$ factorization, the topological susceptibility, the glueball, meson and baryon spectra, the chiral dependence of masses and decay constants, and weak matrix elements related to  the $\Delta I=1/2$ rule in kaon decays.
\PACS{
{11.15.Pg,12.38.Gc,11.15.Ha,13.25.Es}{}
     } 
} 

\maketitle

\section{Introduction}
\label{sec:intro}

It is well known that $SU(N_c)$ gauge theories, with and without matter content, simplify in the so called 't Hooft or planar limit  \cite{tHooft:1973alw}, which corresponds to 
\begin{eqnarray}
N_c\rightarrow\infty,\;\;\; \lambda=g^2 N_c ={\rm fixed} ,
\end{eqnarray}
where $g$ is the standard gauge coupling.  This limit is non-trivial, as can be expected from the fact that asymptotic freedom survives. This may be seen in the running of the 't Hooft coupling,
\begin{eqnarray}
\mu {d \lambda \over d \mu} = - {11\over 3} {\lambda^2\over  8 \pi^2} +{\mathcal O}(\lambda^3),
\end{eqnarray}
which implies that a non-perturbative scale, $\Lambda^\infty_{\rm QCD}$, will be generated dynamically. The growth of the coupling at low energies results in charge
confinement and spontaneous chiral symmetry breaking. The 't Hooft limit therefore captures the most important non-perturbative phenomena of QCD.

In spite of the increased number of degrees of freedom, the theory simplifies to the extent that 
precise non-perturbative predictions can be made. In fact, it has been a long-term aspiration that the theory could be solved  in this limit. An example of the simplification of the large $N_c$ limit is the remarkable Eguchi-Kawai (EK) reduction \cite{Eguchi:1982nm}, which shows that finite volume effects are absent. Under certain conditions, the $SU(\infty)$  Yang-Mills theory can even be reduced to a matrix model on a single site.  Even though the planar limit of QCD has not been solved analytically, it is unquestionable that the large $N_c$ limit has been of great importance in the understanding of QCD, both from a theoretical as well as phenomenological viewpoint. 

Perturbative arguments indicate that the planar limit of QCD is a theory of free and infinitely narrow glueballs and hadrons \cite{tHooft:1973alw,Witten:1979kh,Coleman:1980nk}. The hope is that the spectrum of planar QCD provides a good approximation to that at $N_c=3$. The description of  interactions and decays requires however non-vanishing  $1/N_c$ corrections. On the other hand, some of the planar predictions  are known to fail dramatically, such as the famous $\Delta I=1/2$ rule in kaon decays, indicating the relevance of at least some of the subleading corrections. Lattice QCD can provide a quantitative, first-principles determination of the subleading ${\mathcal O}(1/N_c)$ corrections to the 't Hooft limit by directly simulating gauge theories at different $N_c$\cite{Teper:1998te}. The strict planar limit can be most efficiently explored by numerical methods using the celebrated EK reduction and its variants. Progress in this area has been recently reviewed in \cite{GarciaPerez:2020gnf}.  

In this review, we will concentrate on the recent lattice explorations of $SU(N_c)$ gauge theories with varying number of colours at zero temperature.  We refer to earlier reviews on the subject  \cite{Panero:2012qx,Lucini:2012gg,Lucini:2014bwa} for further results.  Although our focus will be QCD, some of the computations reviewed might be interesting in the context of compositeness models of electroweak symmetry breaking \cite{DeGrand:2019vbx,Drach:2020qpj} and dark matter models  \cite{Kribs:2016cew}. 

The structure of the review is as follows. In sec.~\ref{sec:largencpheno}, we review the main large $N_c$ predictions that have been tested on the lattice.  In sec.~\ref{sec:largenclattice},  we  discuss the lattice approaches to the planar limit and the important issue of scale setting in this context. The main results in Yang-Mills at varying $N_c$ are described in sec.~\ref{sec:yangmills}, while hadronic observables are discussed in sec.~\ref{sec:hadrons}. We end with some concluding remarks.

\section{Non-perturbative predictions at large $N_c$}
\label{sec:largencpheno}

\begin{figure}[ht]
  \centering
  \includegraphics[width=0.8\linewidth]{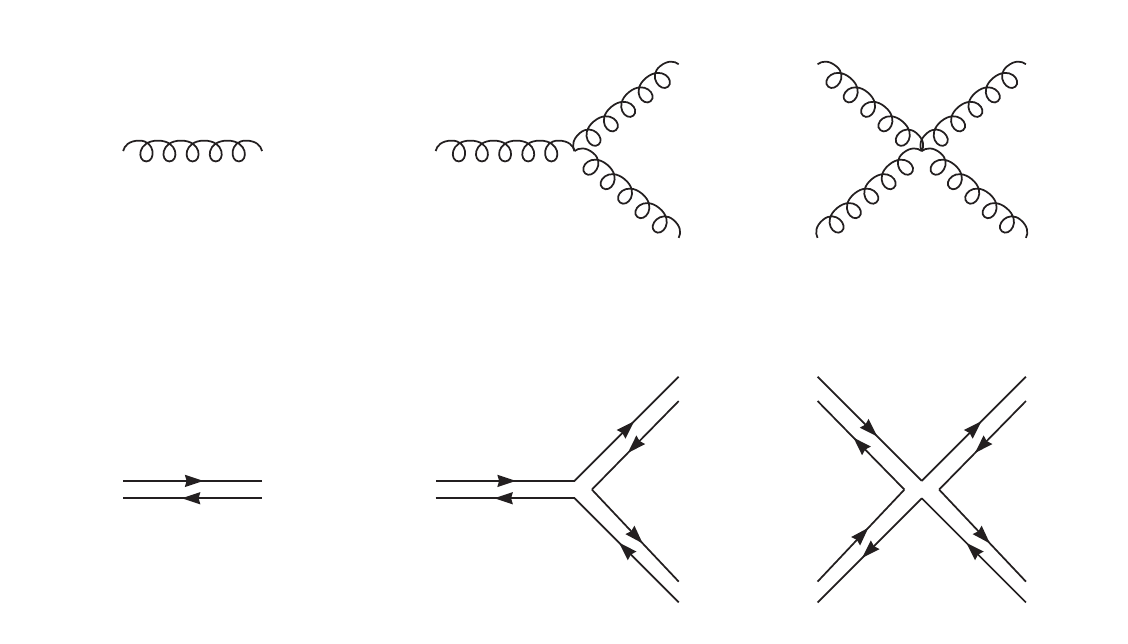}  
  \caption{'t Hooft double-line notation.}
  \label{fig:doubleline}
\end{figure}

In the absence of an analytical solution to the planar limit of Yang-Mills theories,  the known predictions are based on counting powers of $N_c$ in perturbation theory \cite{tHooft:1973alw}.  A quark or antiquark is in the  $N_c$ or $\overline{N}_c$ representation of  $SU(N_c)$ and has one colour index, it is therefore represented by a single directed line. A gluon is in the adjoint representation, which can be constructed out of the product of  an $N_c$ and an $\overline{N}_c$, and can therefore be conveniently represented as  a double line with opposite directions. The usual Feynman notation for gluons is translated to the 't Hooft notation as depicted in Fig.~\ref{fig:doubleline}. 
It is easy to see that planar vacuum diagrams involving only gluons scale as $N_c^2$ to all orders in the loop expansion, while  diagrams with one fermion loop at the boundary and any internal planar exchange of gluons scale as $N_c$. This result can be derived, noting that a generic Feynman diagram in the double-line notation looks like polygons attached by the double lines to patch a surface. For planar diagrams the surface has the topology of a sphere, while non-planar diagrams contain handles, and fermion loops add holes. The power counting in $N_c$ can be shown to be related to the topology of the surface, more concretely to its Euler character, $\chi$. A diagram with $h$ handles and $f$ holes scales as
\begin{eqnarray}
N_c^\chi = N_c^{2-2 h - f}. 
\end{eqnarray}

The first prediction of the large $N_c$ limit is that, if the theory is confining,  it must be a non-interacting theory with an infinite tower of stable particles \cite{Witten:1979kh,Coleman:1980nk}.  To see this, let us consider a singlet gluonic operator such as
\begin{eqnarray}
O_G(x) = {1\over N_c} {\rm Tr}[F_{\mu\nu}^2].
\end{eqnarray}
Diagrammatically, the planar contributions to the connected $n$-point correlation function scale as
\begin{eqnarray}
\langle O_G(x_1)... O_G(x_n) \rangle_c& \sim & N_c^{2-n}.
\end{eqnarray}
Let us assume that there is a pole in the two point function, that is a particle in the spectrum, $|p\rangle$, the scaling of the two-point function implies
\begin{eqnarray}
\langle 0 | O_G(x) | p\rangle \sim N_c^0,
\end{eqnarray}
and consequently the decay amplitude, that can be obtained from the three-point function, is suppressed by $N_c^{-1}$, while the scattering amplitude $2\rightarrow 2$ from the four-point function is suppressed by $N_c^{-2}$. The particle does not decay nor scatter as $N_c\rightarrow \infty$. 

In addition, the number of one particle states cannot be finite, because this would imply that the large momentum behaviour of the two point function
\begin{eqnarray}
\int d^4 x e^{-i p x} \langle O_G(x) O_G(0) \rangle_c  =\sum_i {g_i^2\over p^2 + m_i^2} \sim {\mathcal O}(p^{-2}) , 
\end{eqnarray}
while asymptotic freedom predicts powers of $\log(p^2/m_i^2)$. This can only happen if the sum over the spectrum  contains an infinite number of states.

\subsection{Mesons}
A similar analysis can be done for mesons. Considering an operator with the quantum numbers of a meson
\begin{eqnarray}
O_M^a(x) = {1\over \sqrt{N_c}} \bar{\psi}(x) T^a \Gamma \psi(x),
\label{eq:normM}
\end{eqnarray}
the large $N_c$ counting gives
\begin{eqnarray}
\langle O_M(x_1)...O_M(x_n)\rangle_c \sim N_c^{1-{n\over 2}},
\label{eq:McorNc}
\end{eqnarray}
and therefore, if there are mesonic states in the spectrum, they are stable and non-interacting in the 't Hooft limit. The operator $O_M$ creates a meson from the vacuum with 
amplitude $N_c^0$, while the connected three-point function scales as $\sim 1/\sqrt{N_c}$, and the four-point function like $\sim N_c^{-1}$. The tower of resonances must also be infinite for the same argument outlined before for the gluons. 

From eqs.~(\ref{eq:normM}) and (\ref{eq:McorNc}), it follows 
\begin{align}
F^2_\pi &\propto  \langle A_\mu^a(x) A_\mu^a(y) \rangle \propto   N_c,\;\;\ \nonumber\\
 \Sigma &= -\langle  \bar{\psi}(x) \psi(x) \rangle \  \propto  N_c,
\end{align}
while the scattering length
\begin{eqnarray}
a_M &\propto &{ |\langle O_M^1 O_M^2 O_M^3 O_M^4 \rangle_c|\over  |\langle 0| O_M |M\rangle|^4} \propto N_c^{-1}.
\label{eq:aM}
\end{eqnarray}

\subsection{Chiral Perturbation Theory}
\label{subsec:chpt}

Spontaneous chiral symmetry breaking survives in the 't Hooft limit~\cite{Coleman:1980mx}, and therefore the lightest states in the spectrum are 
the light pseudoscalar mesons. Their mass is ${\mathcal O}(N_c^0)$, assuming the quark 
mass does not scale with $N_c$, since to leading order in the quark mass:
\begin{eqnarray}
M^2_\pi =  {\Sigma\over F_\pi^2} (m_u+ m_d).
\end{eqnarray}
Chiral perturbation theory (ChPT) and the chiral Lagrangian  \cite{Gasser:1984gg}
describe accurately meson interactions at low energies as an expansion in momentum and
quark mass:
\begin{eqnarray}
{\mathcal L}_{\chi} = {\mathcal L}^{(2)}_{\chi}+ {\mathcal L}^{(4)}_{\chi}+...,
\end{eqnarray}
with 
\begin{eqnarray}
{\mathcal L}^{(2)}_{\chi} &=& {F^2 \over 4} {\rm Tr}[\partial_\mu U^\dagger \partial^\mu U] -{\Sigma \over 2} {\rm Tr}[M_q U +h.c.],\nonumber\\
{\mathcal L}^{(4)}_{\chi} &=& \sum_{i=1,10} L_i {\mathcal O}_i[U,\partial_\mu, M_q],
\end{eqnarray}
where $M_q$ is the quark mass matrix, $U= \exp( i {\sqrt{2} \pi\over F})$ and  $\pi$ represents the pions, in the adjoint representation of $SU(N_f)$. The operators ${\mathcal O}_i$ are  Lorentz and chirally invariant  combinations of $U$, with four derivatives or two quark mass matrices. The $L_i$ are the famous low-energy couplings (LECs)~\cite{Gasser:1984gg}.
The expansion parameter in ChPT is
\begin{eqnarray}
   \left(M_\pi\over 4 \pi F_\pi\right)^2 \sim  \left(p \over 4 \pi F_\pi\right)^2  \sim  {1\over N_c},
\end{eqnarray}
and becomes smaller and smaller as $N_c \rightarrow \infty$. However, the range of validity to the chiral effective theory does not  increase: the failure of the chiral expansion will be abrupt when the energy reaches the mass of the heavy resonances, $\Lambda_\chi$. These masses are expected to scale as $N_c^0$, and therefore remain constant as we approach $N_c\rightarrow \infty$.  One may typically consider $\Lambda_\chi \sim M_\rho$. 

  An additional simplification of the chiral Lagrangian comes from the fact that only a subset of the low-energy couplings \cite{Gasser:1984gg} are leading in $N_c$, ie. $L_i \propto {\mathcal O}(N_c)$. These are the ones that correspond to chiral operators (${\mathcal O}_i$) with a single flavour trace. This is because according to the general rules, diagrams with more than one fermion loop are subleading. More concretely, for $N_f=3$ \cite{Gasser:1984gg,Peris:1994dh}:
\begin{eqnarray}
L_1,L_2,L_3,L_5,L_8,L_9,L_{10} &\propto & {\mathcal O}(N_c), \nonumber\\
2 L_1-L_2, L_4,L_6,L_7 &\propto & {\mathcal O}(1).
\end{eqnarray}
Phenomenological approaches have estimated the leading $N_c$ behaviour of these couplings by assuming that the chiral theory matches onto a theory of free resonances, the resonant chiral theory \cite{Ecker:1988te}. The low-energy couplings result from the exchange of heavier resonances and can be extracted in terms of the measured spectrum, and imposing the correct large momentum behaviour of certain correlation functions. For a review see \cite{Pich:2002xy}. 
  
It is well-known that the standard chiral Langrangian needs to be extended to include the $\eta'$ \cite{DiVecchia:1980yfw,PhysRevD.21.3388,Witten:1980sp,Kawarabayashi:1980dp,Gasser:1984gg,Leutwyler:1996sa,HerreraSiklody:1996pm,Kaiser:2000gs}, since the singlet meson becomes degenerate with the remaining pseudoscalar mesons in the planar limit. 
The $\eta'$ mass receives a contribution from the $U_A(1)$ anomaly, that scales as $N_c^{-1}$. 
The Witten-Veneziano (WV) formula \cite{Witten:1979vv,Veneziano:1979ec} (see also \cite{Giusti:2001xh,Seiler:2001je,Giusti:2004qd,Luscher:2004fu,DelDebbio:2004ns}) relates this contribution with the topological susceptibility computed in a theory without quarks, ie. in Yang-Mills (YM):
\begin{eqnarray}
M_{\eta'}^2  \equiv {2 N_f \over F^2_\pi} \int d^4 x \langle q(x) q(0)\rangle_{YM}  \left[1+ {\mathcal O}(N_c^{-1})\right],
\label{eq:W}
\end{eqnarray}
where 
\begin{eqnarray}
q(x) \equiv {\lambda \over 32 \pi^2 N_c} {\rm Tr}[F_{\mu\nu}(x) {\tilde F}^{\mu\nu}(x)] ,
\end{eqnarray}
and we have considered massless quarks. The topological susceptibility measures the $\theta$ dependence of the vacuum energy, which should vanish for massless quarks. The WV  relation can be deduced  from requiring that the $\eta'$ contribution to the $\theta$ dependence exactly cancels the leading  gluonic contribution, or from the anomalous chiral Ward identity.  This is one example where the diagrammatic analysis leads to the wrong conclusion: the leading $N_c$ scaling of the  topological susceptibility is cancelled by a  naively subleading one. 

The presence of the $\eta'$ in the chiral Lagrangian makes it necessary to correlate the large $N_c$ and  the momentum/mass expansions. The pion manifold is now $U(N_f)$, and a consistent power counting  \cite{HerreraSiklody:1996pm,Kaiser:2000gs}  ensures the same scaling of the $\eta'$ mass and the remaining pseudoscalar mesons:
\begin{eqnarray}
{\mathcal O}(M_\pi^2) \sim {\mathcal O}(M_{\eta'}^2) \sim {\mathcal O}\left({1\over N_c}\right).
\end{eqnarray}

\subsection{Baryons}

Baryons are trickier in the 't Hooft limit, since their mass has a non-trivial scaling. Nevertheless their properties can also be understood in a systematic expansion in $1/N_c$ \cite{Witten:1979kh,Dashen:1993jt,Jenkins:1993zu,Carone:1993dz,Dashen:1994qi,Dai:1995zg}. By simple arguments, it can be seen that the baryon mass scales with $N_c$. We recall that in QCD a colour singlet can be formed by combining three quarks in the fundamental representation: $\mathbf 3 \otimes\mathbf 3 \otimes \mathbf 3 \to \mathbf 1$. In a general $SU(N_c)$ theory, one would however need $N_c$ fundamental quarks to create a baryon\footnote{An alternative would be including fermions in the antisymmetric representation of $SU(N_c)$, so that large-$N_c$ baryons may still be formed by three fermions \cite{Corrigan:1979xf}.}. This motivates the simple expectation for the baryon mass: $M_B = m_0 N_c + O(N_c^0)$. Moreover, the baryon--baryon to n-meson amplitude scales like $\sqrt{N_c}^{1-n}$. 

Spin-flavour symmetry results in  consistency conditions that  allow to understand for example the baryon hyperfine splitting \cite{Jenkins:1993zu}:
 \begin{eqnarray}
{ M_B\over N_c} =  m_0+ B {J (J+1)\over N_c^2} + C \left({J (J+1)\over N_c^2}\right)^2+...,
 \end{eqnarray}
where $J$ is the baryon spin, and $m_0, B$ and $C$ are $O(N_c^0)$ constants.  For a detailed review on baryons in large $N_c$ 
 see \cite{Manohar:1998xv}.

\subsection{Exotics}
An interesting and timely question is that of exotic states, such as tetraquarks.  The standard lore  \cite{Witten:1979kh,Coleman:1980nk} used to be that tetraquarks cannot exist in 
the 't Hooft limit. The argument however has been revised recently\footnote{Note the mismatch between the arXiv and published versions of Ref. \cite{PhysRevLett.110.261601}. Here we refer to the latter. } \cite{PhysRevLett.110.261601, PhysRevD.88.036016}. The most recent conclusion  is that the large $N_c$ scaling does not allow to establish whether these particles exist or not in the spectrum, however if they do they must be infinitely narrow, just like glueballs and mesons, for the same reasons outlined above. An analysis of the expected scaling of the decay width of tetraquarks with different flavour content can be found in\cite{PhysRevD.88.036016}.  

 It is interesting to analyse the old argument supporting the non-existence of exotics \cite{Witten:1979kh,Coleman:1980nk}. It relies on the fact that the two point correlation function of two tetraquark operators (constructed from two quark bilinears) has a factorized contribution, leading in $N_c$, corresponding to the propagation of two non-interacting mesons. However,  the tetraquark pole, if it exists, should appear in the connected part of the correlator\cite{PhysRevLett.110.261601} which is subleading in $N_c$. In other words the factorized and connected terms might describe the dynamics of different processes, one is not necessarily the subleading $N_c$ correction of the other, and both can in principle survive the 't Hooft limit. It might be a challenge in practice to isolate such exotic states from a two meson state, since  the overlap of any operator with the right quantum numbers with the two meson state dominates over that with the exotic one.

\subsection{Factorization}

More generally, large $N_c$ implies the factorization of expectation values of products of singlet operators {in the pure gauge theory}:
\begin{eqnarray}
\langle O_1 O_2 \rangle = \langle O_1 \rangle \langle O_2 \rangle\left[ 1 + {\mathcal O}\left({1\over N^2_c}\right) \right].
\label{eq:fact}
\end{eqnarray}
This property underlies for example the OZI rule and the EK reduction. It should be stressed however that the disconnected and connected parts of any such 
observable could represent different physics.

\subsection{Weak interactions}

Finally, we discuss hadronic weak interactions in the planar limit. One of the most famous failures of the large $N_c$ approximation is the $\Delta I=1/2$ rule, that is the large hierarchy observed in the kaon  decay amplitudes to two pions in the two possible states of isospin $I=0,2$:
\begin{eqnarray}
T[K^0 \rightarrow (\pi\pi)_I ] = i A_I e^{i\delta_I}, \;\;\;\; \left|{A_0\over A_2}\right| \sim 22.
\end{eqnarray}
These amplitudes were established as benchmark lattice calculations since the early days \cite{Brower:1984ta}, but remain very challenging. Only very recently\cite{Boyle:2012ys,Bai:2015nea,Blum:2015ywa,Ishizuka:2018qbn,Abbott:2020hxn}, convincing evidence has been presented that the $\Delta I=1/2$ rule is reproduced near the physical point of $2+1$ simulations.  It is well known that the $\Delta I=1/2$ rule plays a key role in the Standard Model prediction of the direct CP violating ratio $\epsilon'/\epsilon$. 

\begin{figure}[h!]
 \begin{center}
\[\begin{array}{cc}
\raisebox{-.45\height}{\includegraphics[scale=1]{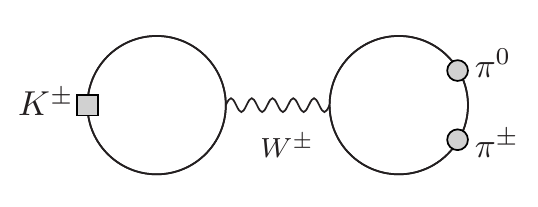}} & \hspace{0.3cm} { \mathcal{O} \Big( N_c^{1/2}\Big)} \\ \\
\raisebox{-.43\height}{\includegraphics[scale=1]{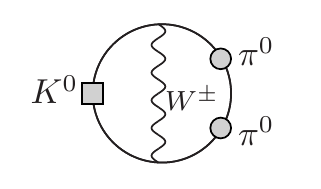}} &\hspace{0.3cm} {\mathcal O}\Big( N_c^{-1/2}\Big)\\ 
\end{array}\]
\caption{\label{fig:deltaI} Leading diagrams in the $N_c$ scaling of  $K^+ \to \pi^+ \pi^0$ (top), and $K^0 \to \pi^0 \pi^0$ (bottom). }
\end{center}
\end{figure}

Surprisingly large $N_c$ misses completely the $\Delta I=1/2$ enhancement \cite{Fukugita:1977df,Chivukula:1986du}. This is easy to see by comparing the large $N_c$ scaling of the decays $K^+\rightarrow \pi^+\pi^0$ and $K^0\rightarrow \pi^0\pi^0$,  which in terms of the isospin amplitudes are:
\begin{eqnarray}
-i T[K^0 \rightarrow \pi^0 \pi^0] &=& \sqrt{2\over 3} A_2 e^{i \delta_2} -{1\over \sqrt{3}} A_0 e^{i \delta_0} \propto {\mathcal O}(N_c^{-{1\over2}}),\nonumber\\
-i T[K^+ \rightarrow \pi^+ \pi^0] &=& {\sqrt{3}\over 2} A_2 e^{i \delta_2} \propto {\mathcal O}(N_c^{1\over 2}),
\label{eq:Ts}
\end{eqnarray}
where the $N_c$ scaling can be easily deduced by inspection of the diagrams, see Fig.~\ref{fig:deltaI}, contributing to each amplitude\footnote{We have included the normalization of each meson state by $\sqrt{N_c}$.}.  This implies that, at the leading order in large $N_c$, the neutral $K^0$ does not decay and therefore from eq.~(\ref{eq:Ts}) it follows: 
\begin{eqnarray}
\left|{A_0 \over  A_2}\right|_{N_c\rightarrow \infty} = \sqrt{2}. 
\end{eqnarray}
One possible explanation of this discrepancy is the fact that the dynamics involves multiple scales: the $W$ mass, the charm mass and hadronic scales, and maybe subleading 
$1/N_c$ corrections could be enhanced by large logarithms.

At energies below the $W$-boson mass, we can integrate it out and represent the weak interactions in the Fermi theory. 
 For example the CP-conserving $\Delta S=1$ transitions the Hamiltonian density takes the simple form:
\begin{eqnarray}
\label{eq:heffs1}
{\mathcal H}^{N_f=4}_{\Delta S=1} =\sqrt{2}G_{\rm F}V_{us}^*V_{ud}   ( k^+ \, \bar{Q}^+(x)+k^- \, \bar{Q}^-(x) )\,, 
\label{eq:hw4}
\end{eqnarray}
with $\bar{Q}^\pm = Z_Q^\pm Q^\pm$ and 
\begin{eqnarray}
Q^\pm =& (\bar{s}_L \gamma_\mu u_L)(\bar{u}_L \gamma^\mu d_L) \pm (\bar{s}_L \gamma_\mu d_L)(\bar{u}_L \gamma^\mu u_L)  \nonumber\\
&-[u\leftrightarrow c].
\label{eq:currento}
\end{eqnarray}
We have neglected quark masses and included all the operators compatible with the ${\rm SU}(4)_{\rm L} \times {\rm SU}(4)_{\rm R}$ flavour symmetries. $Q^+$ transforms as the $(84,1)$, while $Q^-$ transforms as the $(20,1)$.  Only $\bar{Q}^+$ contributes to the $A_2$ amplitude, and therefore a hierarchy in the matrix elements of the two operators is needed to explain the rule. This effective Hamiltonian should be valid down to energies above the charm mass. 
The Wilson coefficients $k^\pm$ do indeed get large logarithms from the $M_W \rightarrow \mu$ \cite{Gaillard:1974nj,Altarelli:1974exa}, but it is not enough to explain the large hierarchy, since $k^-(m_c)/k^+(m_c) \sim 2$. 

Below the charm threshold 
the theory matches onto the  $N_f=3$ theory, that has the form \cite{Buchalla:1995vs}
\begin{eqnarray}
\label{eq:heffs2}
{\mathcal H}^{N_f=3}_{\Delta S=1} =\sqrt{2}G_{\rm F}V_{us}^*V_{ud} \sum_{\sigma=1,...6} k_\sigma Q_\sigma, 
\label{eq:hw3}
\end{eqnarray}
where the operators in the list include all operators that transform as $(27,1)$ or $(8,1)$,  ignoring QED effects. Among the latter, we find the famous penguin operators such as 
\begin{eqnarray}
Q_6 =  (\bar{s}_L \gamma_\mu d_L)\sum_{q=u,d,s} (\bar{q}_R \gamma^\mu q_R).
\end{eqnarray}
The standard lore for a long time was that the matrix elements of the penguin operators were enhanced and, since they are octets that do not contribute to  $A_2$, this fact could be at the origin of the $\Delta I=1/2$ \cite{Vainshtein:1976nd}. This hypothesis does not seem to be confirmed by the last lattice results, where the current operators of eq.~(\ref{eq:currento}) give the dominant contribution to the real part of the isospin amplitudes, as anticipated long ago in Ref. \cite{Bardeen:1986uz}. Instead it has been argued that the enhancement originates in a strong cancellation of the colour-connected and colour-disconnected contributions of the current operator matrix elements \cite{Boyle:2012ys}. Such contributions have a different $N_c$ scaling, and therefore indicate anomalously large $N_c$ corrections. 

At sufficiently low energies and quark masses, a further simplification is possible since the effective hamiltonians, eqs.~(\ref{eq:hw4}) or (\ref{eq:hw3}) can be matched to the corresponding $N_f=4$ (if the charm is considered light) or $N_f=3$ (for a heavy charm) chiral effective theories.  At the leading order of the chiral expansion, the structure of the chiral weak hamiltonian in the $N_f=4$ theory is \cite{Giusti:2004an,Giusti:2006mh}
\begin{eqnarray}
\label{eq:heffs3}
{\mathcal H}^{N_f=4}_{\rm ChPT} =\sqrt{2}G_{\rm F}V_{us}^*V_{ud}   ( g^+ \, {\mathcal Q}^++g^- \, {\mathcal Q}^-)\,, 
\label{eq:chpt4}
\end{eqnarray}
with
\begin{eqnarray}
{\mathcal Q}^\pm &=& {F_\pi^4\over 4} \left[(U \partial_\mu U^\dagger)_{us} (U \partial_\mu U^\dagger)_{du}\pm (U \partial_\mu U^\dagger)_{ds} (U \partial_\mu U^\dagger)_{uu} \right. \nonumber\\
&-&\left. (u\rightarrow c)\right] .
\label{eq:currentochpt}
\end{eqnarray}
It is straightforward to evaluate the ratio of amplitudes in terms of the low-energy couplings $g^\pm$ to lowest order in ChPT:
\begin{eqnarray}
{A_0\over A_2} ={1\over 2 \sqrt{2}} \left(1+ 3 {g^-\over g^+}\right). \label{eq:A0overA2}
\end{eqnarray}
A large ratio $g^-/g^+$ could explain the $\Delta I=1/2$ rule, an effect that would be unrelated to the charm threshold.

These low-energy couplings can be determined from matching appropriate correlation functions in ChPT and lattice QCD\cite{Giusti:2004an,Giusti:2006mh}. The simplest choices are the ratios:
\begin{eqnarray}
R^\pm  = \kern-1.0em
\lim_{ \substack{z_0-x_0\to\infty \\ y_0-z_0\to \infty}}
\frac{\sum_{{\mathbf x},{\mathbf y}} \langle P(y) Q^\sigma(z) P(x) \rangle}
{\sum_{{\mathbf x,\mathbf y}}  \langle P(y) A_0(z) \rangle  \langle P(x) A_0(z) \rangle}, \label{eq:ratios}
\end{eqnarray}
with $A_0(x) \equiv \bar{\psi}^i(x)\gamma_0 \gamma_5 \psi^j(x)$ and $P(x)= \bar{\psi}^i(x) \gamma_5 \psi^j(x)$, with appropriately chosen flavours $i,j$. 

Including the renormalization factors and the Wilson coefficients, we get the physical amplitudes, $A^\pm$,  that  can be matched to the corresponding observables in the chiral theory\cite{Giusti:2004an,Giusti:2006mh}. At lowest order the matching condition is 
\begin{eqnarray}
\lim_{M_\pi \rightarrow 0} A^\pm = g^\pm. 
\end{eqnarray}

\begin{figure}[h!]
 \begin{center}
\renewcommand{\arraystretch}{2}
\[\begin{array}{lll}
 \raisebox{-.5\height}{\includegraphics[scale=0.8]{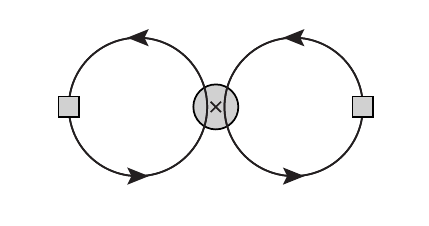}} & \mp &\raisebox{-.5\height}{\includegraphics[scale=0.8]{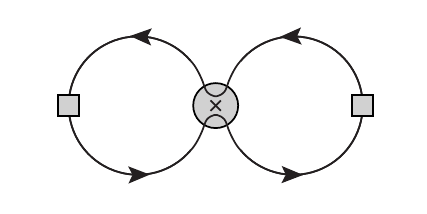}}\\
 \end{array}\]
 \caption{\label{fig:condis} Colour-disconnected (left) and colour-connected (right) contributions to the three-point function in $R^\pm$. }
\end{center}
\end{figure}

\begin{figure}[h]
 \begin{center}
\renewcommand{\arraystretch}{2}
\[\begin{array}{ccc}

{\tilde c}: & \raisebox{-.5\height}{\includegraphics[scale=0.7]{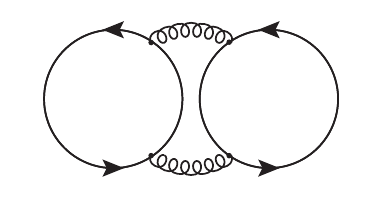}} &{\mathcal O}(N_c^0) \\
{\tilde d}: & \raisebox{-.5\height}{\includegraphics[scale=0.7]{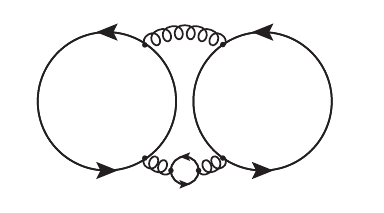}} & {\mathcal O}\left({N_f\over N_c}\right)\\
\end{array}\]
\caption{\label{fig:disnlo} $N_c, N_f$ scaling of various contributions to the colour-disconnected contraction of the three-point function.}
\renewcommand{\arraystretch}{2}
\[\begin{array}{ccc}
\tilde a:   & \raisebox{-.5\height}{\includegraphics[scale=0.7]{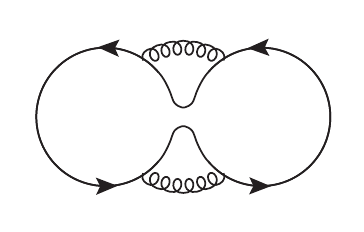}} & {\mathcal{O}}(N_c)\\
\tilde b:&  \raisebox{-.5\height}{\includegraphics[scale=0.7]{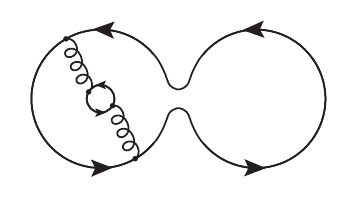}} & \mathcal{O}\left( N_f \right)\\
\end{array}\]
\caption{\label{fig:connlo} $N_c, N_f$ scaling of various contributions to the colour-connected contraction of the three-point function. }
\end{center}
\end{figure}

The $\pm$ three-point functions are the sum of a colour-disconnected diagram and a colour-connected one, that scale differently with $N_c$ , see Fig.~\ref{fig:condis}.
 The relative sign is $\mp$ in the $A^\pm$ amplitudes.
The standard diagrammatic $N_c$ counting results in the following scaling:
\begin{eqnarray}
A^\pm = 1 \pm \tilde a {1 \over N_c}\pm \tilde b {N_f \over N^2_c}+\tilde c {1 \over N^2_c}+ \tilde d {N_f \over N^3_c}+\cdots, \label{eq:rnc}
\end{eqnarray}
where the coefficients $\tilde a- \tilde d$ are 
independent of $N_c$ and $N_f$, and  a {\it natural} expectation is that they are ${\mathcal O}(1)$. Diagrams\footnote{Additional planar exchanges of gluons being of the same order contribute in all cases.} contributing to ${\tilde c}$ and ${\tilde d}$ are
shown in Fig.~\ref{fig:disnlo} while those contributing to $\tilde a$ and $\tilde b$ are in Fig.~\ref{fig:connlo}.

We find that the leading correction, ${\mathcal O}(1/N_c)$ together with the leading effects of dynamical quarks,  ${\mathcal O}(N_f/N_c^2)$, are fully anticorrelated in the ratios.
 This analysis does not predict the sign of the  different terms, i.e., the sign of the $\tilde a-\tilde d$ coefficients, only the (anti)-correlation between the two amplitudes. In particular,  a negative sign of $\tilde a$ and $\tilde b$ results in an enhancement of the ratio $A^-/A^+$.  The scaling of eq.~(\ref{eq:rnc}) is valid for all quark masses, but the different coefficients could be have a different quark mass dependence. 

The lattice can in principle measure all these coefficients and determine whether they have the natural size, ${\mathcal O}(1)$, and quantify to what extent the anti-correlation is enough to explain the $\Delta I=1/2$ rule.

A similar matching to chiral perturbation theory could also be done in the $N_f=3$ flavour theory, where the leading order weak Hamiltonian has also two operators with the same chiral structure, corresponding  to the $27$-plet and octet operators\cite{Bernard:1985wf}. The case of a heavy charm is however much more demanding, not only the list of operators to consider on the lattice is much longer, but it involves the computation of all-to-all propagators. 

There is a long list of phenomenological analyses that have used large-$N_c$ inspired methods to estimate the size of these couplings, in an analogous way to the strong couplings, $L_i$.  Examples are the dual QCD approach \cite{Bardeen:1986vp,Bardeen:1986uz,Buras:2014maa}, the resonance chiral theory \cite{Ecker:1988te} or the chiral quark model \cite{Antonelli:1995nv}, etc. In all cases, the relevance of subleading corrections in $N_c^{-1}$ has been demonstrated, and argued to be related to a strong scale dependence. Incorporating these systematically is not easy and the error resulting from these approaches is therefore quite uncertain. For a review see \cite{Cirigliano:2011ny}.

\section{Lattice QCD at large $N_c$}
\label{sec:largenclattice}

Two approaches have been followed to explore the 't Hooft limit in the lattice formulation. The most straightforward approach is to simulate\footnote{HiRep is a publicly available lattice QCD code that allows one to simulate generic $SU(N_c)$ theories with different fermion content  \cite{DelDebbio:2008zf,DelDebbio:2009fd}.} at different values of $N_c$  \cite{Teper:1998te,Lucini:2001ej,Lucini:2003zr,Lucini:2008vi}, and study the scaling of any physical quantity in units of some reference scale, such as $\Lambda_{QCD}$, which is well defined when $N_c\rightarrow \infty$. {The cost of the simulation increases with $N_c$, in particular, for dynamical simulations the cost of a single configuration scales like $\sim N_c^2$, dominated by matrix-vector multiplications for the inversion of the Dirac operator. However, statistical fluctuations seem to decrease, as can be expected from the factorization property, eq.~(\ref{eq:fact}). This was also observed in Refs. \cite{Hernandez:2019qed,Donini:2020qfu}, although a large $N_c$ scaling of the autocorrelation time has not been done yet.} Reaching very large values of $N_c= {\mathcal O}(100)$ is however not possible with this method. The Eguchi-Kawai reduction\cite{Eguchi:1982nm} allows to study the leading behaviour in $N_c$ much more efficiently by exploiting the volume independence ensured by factorization, and reducing the model to a single-point lattice model. Different variants of the EK reduction exist \cite{BHANOT198247,Narayanan:2003fc,Kovtun:2007py,Unsal:2008ch,GONZALEZARROYO1983174,PhysRevD.27.2397} in the literature, and they have been successfully implemented to compute the 't Hooft limit of several physical observables \cite{Gonzalez-Arroyo:2014dua,GONZALEZARROYO1983415,FABRICIUS1984293,GonzalezArroyo:2012fx,Gonzalez-Arroyo:2015bya,Hietanen:2009tu}. We will show some of the most recent results, but a more detail account of reduced models can be found in \cite{GarciaPerez:2020gnf}. 

The complementarity of both approaches is evident when trying to evaluate the subleading $1/N_c$ corrections. While these corrections cannot be captured by the reduced models, they 
can be straightforwardly computed with the direct method. The input provided by the reduced model on the planar limit is nevertheless very helpful to constrain the asymptotic $N_c$ dependence.

\subsection{Scale setting and $N_c$-dependence}
\label{sec:scale}

The outcome of a lattice calculation is always a either dimensionless ratio, or a dimensionful quantity in terms of the lattice spacing, $a$. In QCD, the value of the lattice spacing is typically fixed by some experimental input (e.g., a mass or decay constant). Once $a$ is known, any observable can be converted to physical units (MeV, fm,...). The procedure of computing $a$ in physical units is known as the (absolute) scale setting. Of course, a complete calculation will require simulations at multiple values of $a$, and a continuum limit extrapolation for the observables of interest.

In the context of large $N_c$, the scale setting is a more subtle issue, and it necessarily involves some arbitrariness. To illustrate this, let us consider the case of a pure $SU(N_c)$ gauge theory.  We could perform simulations with varying $N_c$, and compute the ratio of two different scales, $\Lambda_1 / \Lambda_2$ in the continuum limit. These scales could for instance be masses of glueballs with different quantum numbers. Then, we could fit the functional form of the $N_c$-dependence as:
\begin{equation}
\frac{\Lambda_2}{\Lambda_1} = c_1 + \frac{c_2}{N_c^2} + \hdots, 
\end{equation}
in the pure gauge theory without fermions.
This is a well-posed problem, and does not require any absolute knowledge of the lattice spacing (only a relative one for the continuum limit). We may however know the value for these two scales from experiments in QCD\footnote{Assuming that the effects of quark loops are negligible.}, and we would like to infer their values as a function of $N_c$. For this, one necessarily needs to impose a known $N_c$ dependence---an all-orders ansatz---for one of the scales of the theory. The simplest choice is to pick, e.g., $\Lambda_1$ as  constant for all values of $N_c$. This fully fixes the prescription for the large $N_c$ limit, and implicitly the lattice spacing at all values of $N_c$ via the scale $\Lambda_1$. It is  clear that different choices of $\Lambda_1$ will produce differences in the subleading $1/N_c$ corrections for other physical quantities, but the same ratios in two different prescriptions will have the same 
$N_c$ scaling. 

From a practical point of view, it is much more convenient to study the $N_c$ dependence by using lattice simulations at a constant line of physics. In other words, to establish the prescription for the large $N_c$ limit at the time of the simulation. For instance, in early work \cite{Lucini:2001ej}, simulations were compared at a constant value of the mean-field improved bare 't Hooft coupling defined as:
\begin{equation}
\lambda_I(a^{-1}) = \frac{2 N_c^2}{\beta P}, 
\label{eq:coupling1}
\end{equation}
where $P$ is the plaquette normalized by $N_c$.
One may say that simulations with constant $\lambda_I$ at different $N_c$ have the same value of the lattice spacing. Thus, after fixing $a$ from $N_c=3$, the large $N_c$ prescription is fully fixed. 

A different line of constant physics was chosen in Ref. \cite{Bali:2013kia}. There, the string tension ---labelled as $\sigma$---was used.\footnote{The string tension, $\sigma$, parametrizes the confining term of the potential for two static quark-antiquark sources, i.e., $V(r) = v + \sigma r + c/r$, with $v,c$ constants.} In this work, the authors used existing results for $\sigma$ as a function of $N_c$ \cite{DelDebbio:2001sj,Lucini:2005vg,Allton:2008ty,Lucini:2012wq,Lohmayer:2012ue} to generate gauge configurations with $N_c \in [2,17]$ and fixed value of $a \sqrt{\sigma}$. Moreover, they used the ``ad hoc value $\sigma = 1$ GeV/fm'' to define their lattice spacing for all values of $N_c$. To further iIlustrate the ambiguity at large $N_c$, we can look at the behaviour of the coupling in eq.~(\ref{eq:coupling1}) in the simulations in Ref. \cite{Bali:2013kia}. We will easily see that $\lambda_I(a^{-1})$ it is not constant, but rather it is modified at order $1/N_c^2$.

In the last few years, the usage of the gradient flow \cite{Luscher:2010iy} has become customary for the scale setting on the lattice \cite{Sommer:2014mea}---also in the context of large $N_c$. The advantage is that the energy density can be used to determine a renormalized 't Hooft coupling in the gradient flow scheme:
\begin{equation}
\langle  E(t) \rangle \equiv \frac{1}{4} \langle G_{\mu \nu }^a G^{\mu \nu}_a \rangle    = \frac{3}{138 \pi^2 t^2} \frac{N_c^2 -1}{N_c} \lambda_{GF},
\end{equation}
with $ \lambda_{GF}$ defined at the scale $1/\sqrt{8 t}$. The gradient flow coupling has been related to the $\overline{MS}$ coupling in perturbation theory \cite{Harlander:2016vzb}. A standard approach in QCD is to define the scale $t_0$ via the energy density as a function of the flow time:
\begin{equation}
 t^2 \langle  E(t) \rangle \big \rvert_{t=t_0} = 0.3. \label{eq:t0}
\end{equation}
Even though $t_0$ cannot be measured experimentally, its value in physical units has been extracted from various lattice calculations \cite{Bruno:2013gha,Sommer:2014mea,Bruno:2016plf}. In Ref. \cite{Ce:2016awn} it was proposed to generalize the definition of $t_0$ to 
\begin{equation}
 t^2 \langle  E(t) \rangle \big \rvert_{t=t_0} = 0.3 \times \frac{3}{8} \frac{N_c^2-1}{N_c},  \label{eq:t0Nc}
\end{equation}
so that results at different $N_c$ may be compared. Moreover, the same authors opt for fixing $\sqrt{t_0} \sim 0.166$ fm for all values of $N_c$.

In dynamical simulations with varying $N_c$, an additional convention\footnote{We assume $N_f$ degenerate flavours for simplicity.} regarding the quark mass must be imposed to compare across values of $N_c$. This is because a generic observable, including $t_0$, depends on the quark masses. An option that fully fixes the convention could be matching the chirally extrapolated value of $t_0$ for different gauge groups. In Ref. \cite{DeGrand:2017gbi} with $N_f=2$ simulations, two different conditions were imposed: (i) the ratio $\sqrt{t'_0}/r_1$ is fixed, where $t'_0$ for $N_c=3$ is the standard one in eq.~(\ref{eq:t0}), and $r_1$ is related\footnote{The modified Sommer parameter, $r_1$, is defined in terms of the force between static quarks via the implicit equation $r^2F(r) = -1.0$ at $r = r_1$. In physical units, $r_1 = 0.31$ fm as measured in real-world QCD \cite{Bazavov:2009bb}.} to the Sommer parameter \cite{Sommer:1993ce,Bernard:2000gd} , and (ii) the ratio of the vector meson mass over the pion mass is matched. More recently, in Ref. \cite{Hernandez:2019qed} with $N_f=4$ simulations, the generalized value of $t_0$ as in eq.~(\ref{eq:t0Nc}) at $M_\pi = 420$ MeV was fixed for all values of $N_c$.

\begin{figure}[h!]
\centering 
\includegraphics[width=.5\textwidth]{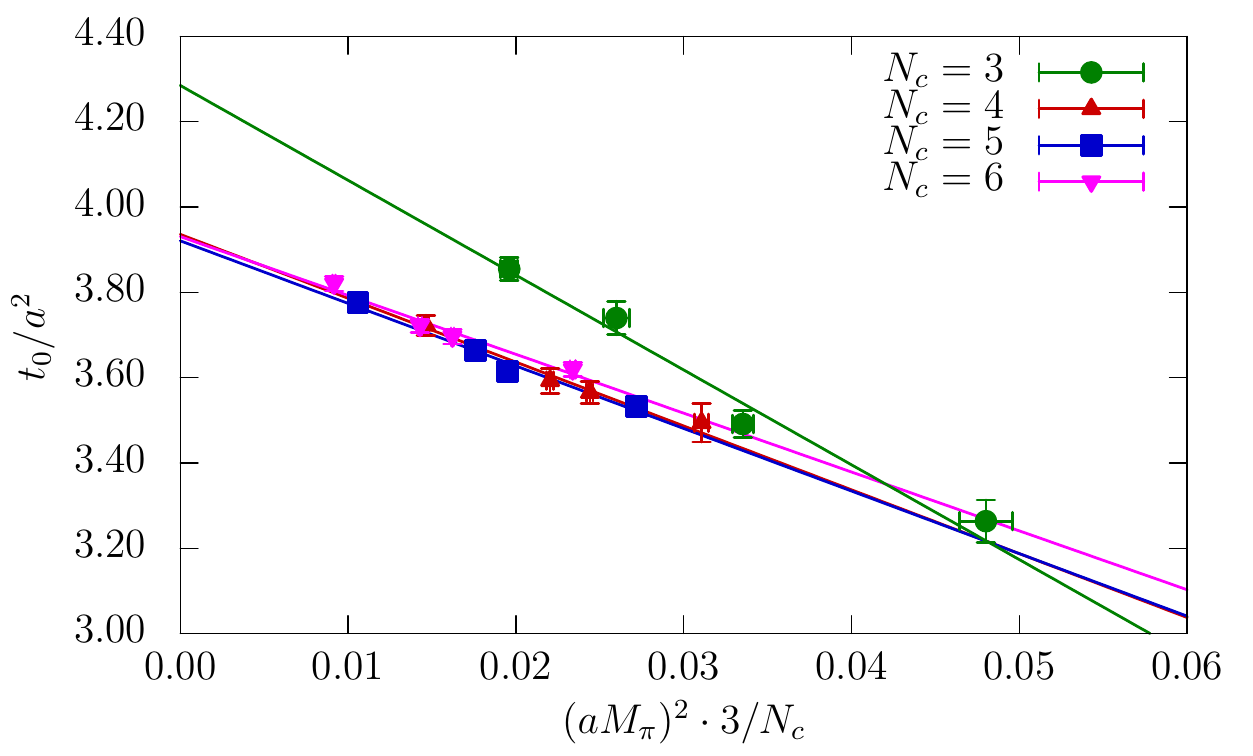}
\caption{Chiral dependence of $t_0$ for various simulations with $N_f=4$ and $N_c=3-6$. Source: Ref. \cite{Hernandez:2019qed}. \label{fig:t0}}
\end{figure}

As a further check of the consistency of this scale setting, one can also study the chiral dependence of $t_0$ when varying the number of colours. The prediction from Chiral Perturbation Theory is \cite{Bar:2013ora}:
\begin{equation}
t_0(M_\pi) = t_0^\chi \left(1 + \frac{k}{N_c} M_\pi^2 \right) + O(M_\pi^4),
\end{equation}
with $ t_0^\chi, k$ being low-energy constants.  We note that the $1/N_c$ suppression of the mass dependence is consistent with the quenched limit, in which $t_0$ is mass independent. In Fig. \ref{fig:t0}, we show this chiral dependence along with the corresponding chiral fits. As can be seen, the data for $N_c=4,5,6$ lie almost on top of some universal line, whereas $N_c=3$ has larger $1/N_c$ effects.


\section{Yang-Mills theories at large $N_c$}
\label{sec:yangmills}
In this section, we review various results for pure $SU(N_c)$ Yang-Mills theories, including the factorization theorem, the glueball spectrum and topological observables.

\subsection{Large $N_c$ factorization}

One of the simplest yet powerful results in the large $N_c$ limit is the factorization of observables, eq.~(\ref{eq:fact}). This result is derived from a perturbative analysis, but can be tested non perturbatively in lattice simulations at different values of $N_c$. Recently, such study  was carried out in Refs. \cite{Vera:2018lnx,GarciaVera:2017xif}.

\begin{figure}[h!]
\centering 
\includegraphics[width=0.5\textwidth]{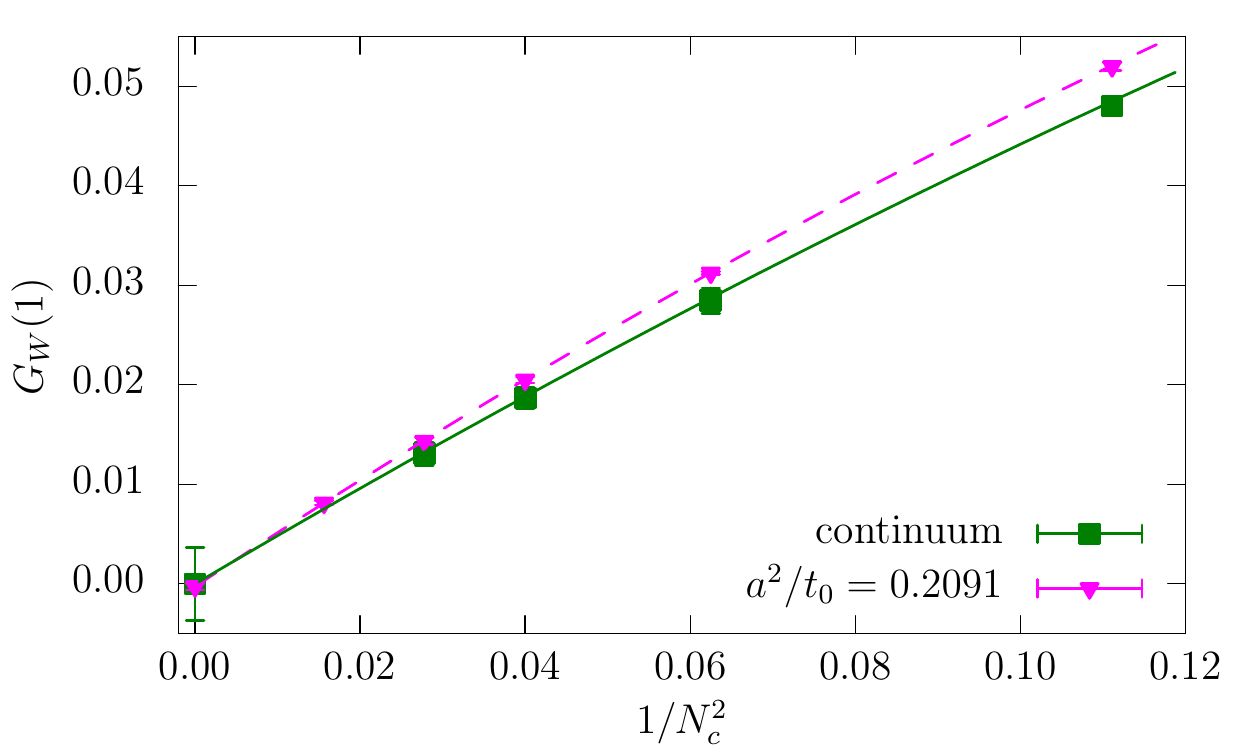}
\caption{ Large $N_c$ extrapolations of $G_W (1)$ in the continuum (solid line) and at finite lattice spacing (dashed line). Source: Ref. \cite{Vera:2018lnx}.  \label{fig:f1}}
\end{figure}

In order to test the factorization hypothesis, the authors use various quantities, e.g., a square spatial Wilson loop at the point $(x_0,{\mathbf x})$ with size $R_c$ and Wilson-flowed to time $t$:
\begin{align}
\begin{split}
W(c) &= \langle W(t, x_0, \vec x, R_c)  \rangle , \\ W^{sq}(c) &= \langle W^2(t, x_0, \vec x, R_c)  \rangle,
\end{split}
\end{align}
The flow time is a fraction of $t_0$, eq.~(\ref{eq:t0Nc}),  $t =c t_0$, and the size is chosen as \footnote{The radius of a square Wilson loop is always an integer. In this work, the authors interpolate between integers using the methods described in Ref. \cite{Vera:2017dxr}.} $R_c = \sqrt{8 c t_0}$. Note that the authors use open boundary conditions, to avoid topology freezing, and $x_0$ must be taken far from the boundary. The ratio
\begin{equation}
G_W(c) = \frac{W^{sq}(c) - W^2(c)}{W^2(c)}, \label{eq:Gw}
\end{equation}
according to the factorization property, 
is expected to be of order $1/N_c^2$. The results for $G_W(1)$ in this study are shown in Fig. \ref{fig:f1}. As can be seen, the authors find very good consistency with the factorization hypothesis both at a finite lattice spacing (fixed $t_0/a^2$), and in the continuum limit. For other gluonic observables, the authors also find the expected large $N_c$ scaling and confirm the factorization hypothesis. 

\begin{figure}[h!]
\centering 
\includegraphics[width=0.5\textwidth]{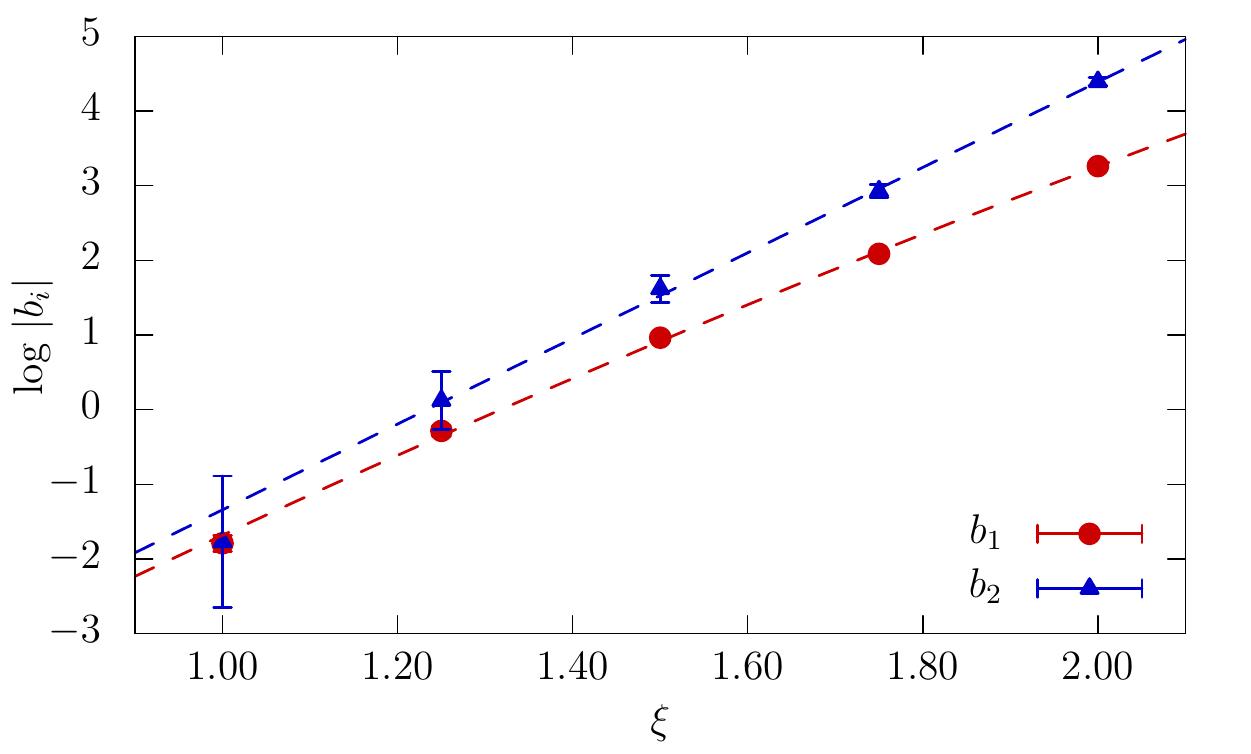}
\caption{ Dependence of the  corrections to factorization with the size of the Wilson loop, $\xi$ [see eq.~(\ref{eq:hatGw})]. The $b_i$ parameters are those of eq.~(\ref{eq:fitGNc}). For each parameter, we include a quadratic interpolating line. Source: Ref. \cite{Vera:2018lnx}.  \label{fig:f2}}
\end{figure}

The dependence of $G_W(c)$ on the size of the loop is also studied. For this, they define a new observable $\hat G_W(\xi)$, that depends on the Wilson loop 
\begin{equation}
\widehat W(\xi) = \langle W(t_0/2, x_0, \vec x, R_\xi)  \rangle   \label{eq:hatGw}
\end{equation}
with varying radius, $R_\xi = \xi \sqrt{4 t_0}$. The authors fit the $1/N_c$ scaling as:
\begin{equation}
\widehat{G}_W(\xi) = b_0 + b_1 \frac{1}{N_c^2} + b_2 \frac{1}{N_c^4}. \label{eq:fitGNc}
\end{equation}
As expected, $b_0$ is always compatible with zero. In contrast, the results for the coefficient of the subleading corrections , shown in Fig. \ref{fig:f2}, seem to grow exponentially with the size of the Wilson loop.

This shows that the coefficient of the subleading corrections can be large, but this is not necessarily a signal of failure of the large $N_c$ expansion. This is because the leading 
and subleading contributions can be related to different physics scales. While the disconnected and leading term is controlled by the static potential, the connected part could get 
contributions from other physics scales.

\subsection{Glueballs at large $N_c$}
\begin{figure*}[t!]
\centering 
\includegraphics[width=1\textwidth]{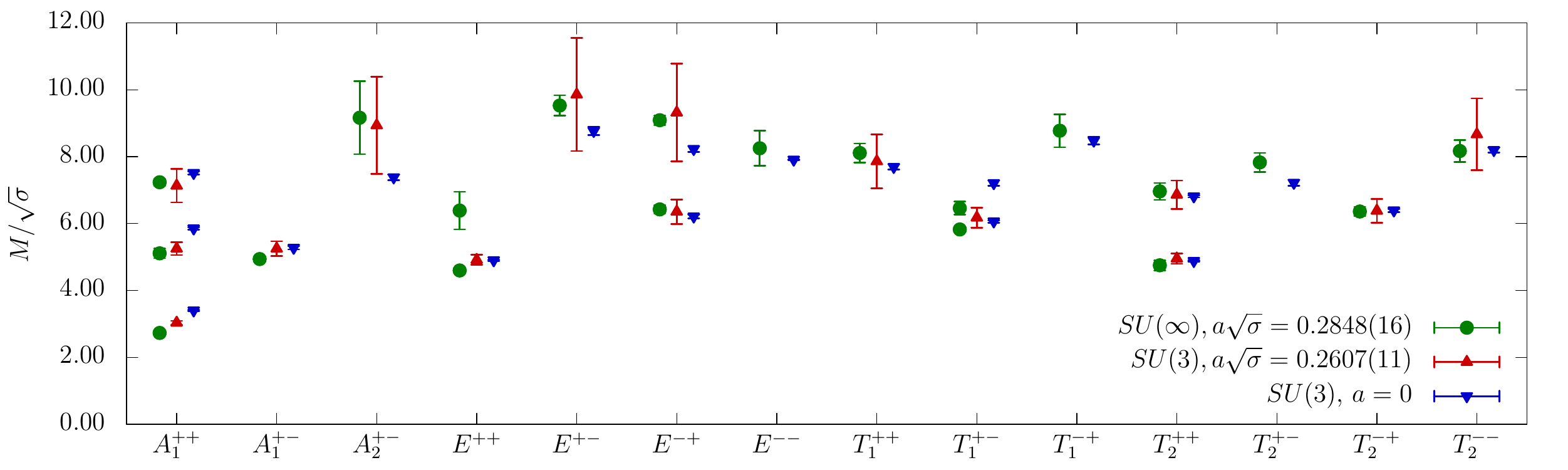}
\caption{ Spectrum of glueballs at finite lattice spacing in the large $N_c$, and $N_c=3$ from Ref. \cite{Lucini:2010nv}. Some of the $SU(3)$ levels were not determined in Ref. \cite{Lucini:2010nv}. As a comparison, we also include the continuum limit results for $SU(3)$ from Ref. \cite{Athenodorou:2020ani}. \label{fig:glueballs}}
\end{figure*}

The glueball spectrum is a very interesting nontrivial manifestation of non perturbative dynamics in the pure Yang-Mills theory. Describing glueballs in the pure gauge theory may help us finding their analogues in full QCD, assuming their mix with $\bar{q} q$ states is not large. Moreover, the lowest spectrum of a pure $SU(N_c)$ that weekly interacts with the SM particles has been proposed as a candidate for dark matter \cite{Soni:2016gzf}---see also Ref. \cite{Kribs:2016cew}.

The dependence of the glueball spectrum on the number of colours has been subject of study for many years \cite{Teper:1998kw,Lucini:2001ej,Lucini:2004my,Meyer:2004jc,Meyer:2004gx,Lucini:2010nv,Amato:2015ipe}, including the $\theta$-dependence\footnote{$\theta$ is the coefficient of the topological term in the Lagrangian.} \cite{DelDebbio:2006yuf}. Early calculations determined the continuum result for the lowest-lying glueballs, with $J^{PC}=0^{++}, 2^{++}$, in the large $N_c$ limit \cite{Lucini:2004my}. Up to date, the most extensive $SU(N_c)$ glueball spectroscopy result is that of Ref. \cite{Lucini:2010nv}, with the caveat that no continuum limit was performed. In Fig. \ref{fig:glueballs}, we compare the large $N_c$ results of Ref. \cite{Lucini:2010nv} to the $N_c=3$ ones of the same work, both at similar value of the lattice spacing. As can be seen, $1/N_c^2$ corrections are quite small, but sometimes resolvable. We also include in the comparison  the continuum limit results for $SU(3)$ in Ref. \cite{Athenodorou:2020ani}. There are signs of significant discretization effects in $SU(3)$ for the states that have been measured with good statistical precision. It would be interesting to compute the large $N_c$ result including a continuum extrapolation.

Some of the states in the low-lying glueball spectrum correspond to two-particle scattering states, as discussed in Ref. \cite{Meyer:2004vr}. These levels have been omitted in Fig. \ref{fig:glueballs}. In the large $N_c$ limit, their energy is simply given by the non-interacting one, as glueball interactions are suppressed with $N_c$. Still, at large but finite $N_c$ these levels can be used to study interactions among glueballs. So far, these studies have only been attempted for $SU(2)$ using the HAL QCD method \cite{Yamanaka:2019gak,Yamanaka:2019yek,Yamanaka:2019aeq}. It would be interesting to perform a conclusive Lüscher analysis of two-glueball interactions, and its $N_c$-dependence.

{It is also worth mentioning that the large $N_c$ limit of $Sp(2N_c)$ and $SU(N_c)$ coincides, as shown in Ref. \cite{Lovelace:1982hz}. Thus, the $Sp(2 N_c)$ results in Refs. \cite{Bennett:2020hqd,Bennett:2020qtj} are also relevant for the large $N_c$ limit of QCD.  A confirmation of the expectation, supported by the data, is that the ratio of masses for the tensor and scalar glueball seems to be independent of the gauge group. In particular, it agrees for $Sp(\infty)$ and $SU(\infty)$.}

\subsection{Topological susceptibility}
\label{subsec:topsus1}
The topological susceptibility is the first momentum of the distribution of the topological charge, $Q$:
\begin{equation}
\chi_{YM} = \lim_{V \to \infty} \frac{1}{V} \langle Q^2 \rangle.
\end{equation}
As we have seen in sec.~\ref{sec:largencpheno}  the  Witten-Veneziano equation, eq.~(\ref{eq:W}), relates this observable in Yang-Mills  to the mass of the $\eta'$ meson in the planar limit of QCD. Studying the size of this subleading $1/N_c$  is important to quantify how far from the planar limit  QCD is.

\begin{figure}[h!]
\centering 
\includegraphics[width=0.5\textwidth]{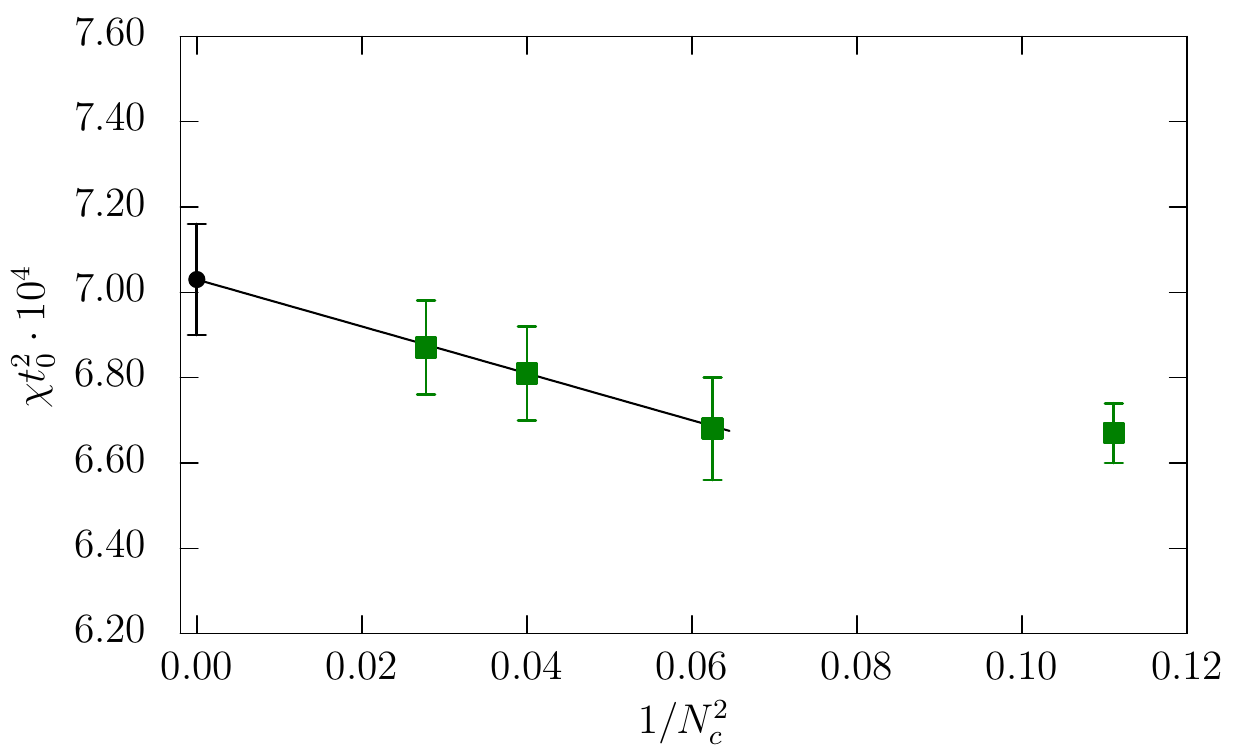}
\caption{Scaling of the topological susceptibility with the number of colours in the pure Yang-Mills theory. The dashed line is a linear fit in $1/N_c^2$ including up to $N_c=4$. Source: Ref. \cite{Ce:2016awn} for $N_c>3$ and Ref. \cite{Ce:2015qha} at $N_c=3$. \label{fig:topsus}}
\end{figure}

The first computation of the topological susceptibility in the large $N_c$ limit was attempted two decades ago, in Ref. \cite{Lucini:2004yh}. One of the main difficulties is the critical slowing down \cite{Schaefer:2010hu}, that is, the very long autocorrelation time in topological observables as one approaches the continuum limit using periodic boundary conditions\footnote{In a recent article \cite{Bonanno:2020hht}, a novel algorithm has been proposed to mitigate this problem.}. This is already present in $N_c=3$, and it becomes more severe for $N_c>3$. 

The use of open boundary conditions \cite{Luscher:2011kk} significantly improves this situation, and has been an essential ingredient in the recent computation of  $\chi_{YM}$ in the continuum limit \cite{Ce:2015qha,Ce:2016awn}. In Fig. \ref{fig:topsus}, we show the continuum result of $\chi_{YM}$ as a function of the number of colours. The large-$N_c$ result reported by the authors is
\begin{equation}
\lim_{N_c \to \infty} t_0^2 \chi_{YM} = 7.03(13) \cdot 10^{-4},
\end{equation}
which differs from the $N_c=3$ result by only about $O(5\%)$. As pointed out by the authors, the small size of $1/N_c^2$ effects explains why the $N_c=3$ result is already able to predict the bulk result of the $\eta'$ mass. A lattice test of the Witten-Veneziano equation, computing directly the $\eta'$ mass, has also been carried out in Ref. \cite{Cichy:2015jra} for $N_c=3$.

The large-$N_c$ topological susceptibity has also been studied through the $\theta$-dependence  in Refs. \cite{DelDebbio:2002xa,Bonati:2016tvi,Kitano:2020mfk}, and in lower dimensional theories  in Refs.~\cite{Bonati:2019ylr,Bonanno:2018xtd}.

\section{Hadronic quantities at large $N_c$  }
\label{sec:hadrons}
Here, we review recent results of hadronic quantities in the context of the large $N_c$ limit of QCD. We start with the meson and baryon spectrum. Then we discuss the chiral dependence of the meson masses, decay constants, scattering lengths and the topological susceptibility. We then turn to novel results on weak amplitudes related to the $\Delta I=1/2$ rule. 

\subsection{Meson spectrum}

The simplest question one may ask regarding hadrons is how the spectrum depends on the  number of colours. As already discussed, all mesons are expected to become stable particles at large $N_c$, and thus they may be treated as asymptotic states, rather than resonances. 

\begin{figure}[h!]
\centering 
\includegraphics[width=0.5\textwidth]{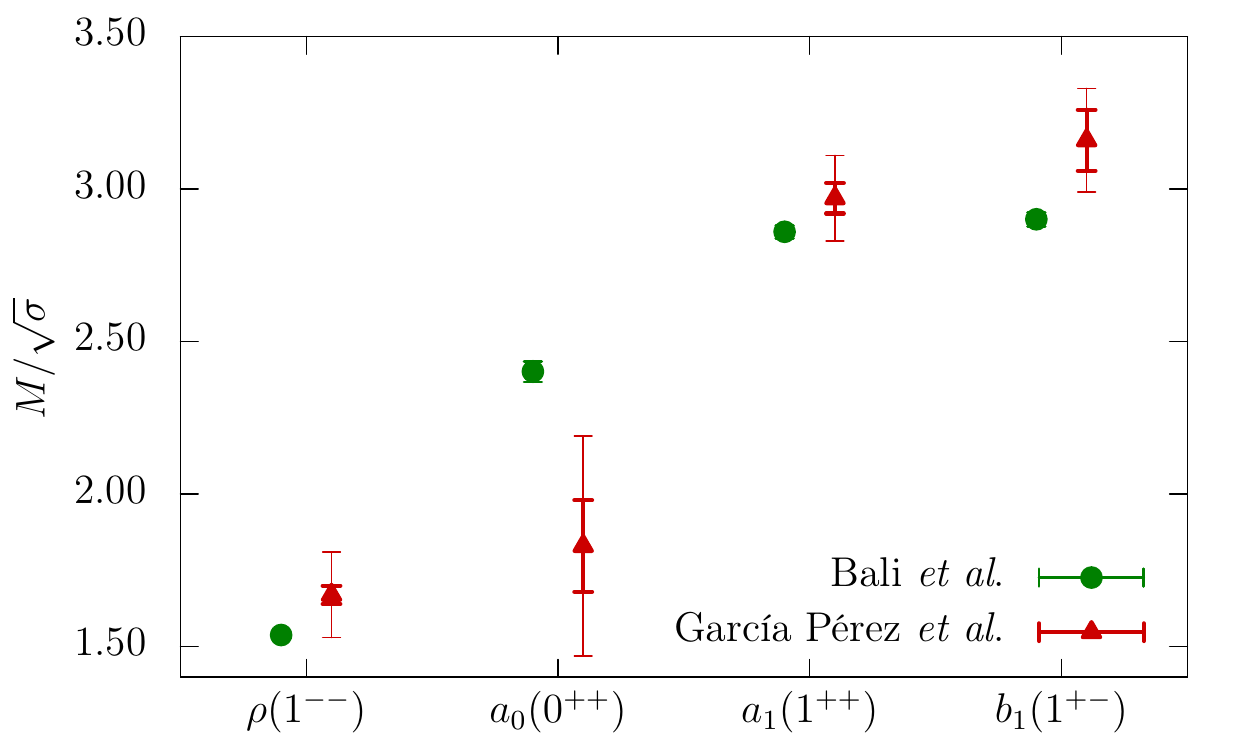}
\caption{Large $N_c$ meson spectrum in different channels, with $J^{PC}$ as provided in the x-labels. We compare the results from standard large-$N_c$ approach from Ref. \cite{Bali:2013kia}, to the ones obtained using the Twisted Eguchi-Kawai reduction~\cite{Perez:2020vbn}. In the case of Ref. \cite{Perez:2020vbn}, the smaller errorbars are statistical, whereas the larger ones include the systematics of chiral and continuum extrapolation. \label{fig:mesons}}
\end{figure}

The most extensive study of the meson spectrum was performed in Ref. \cite{Bali:2013kia}. It was done using the quenched approximation at a fixed lattice spacing. Although it is known that quenching alters some $1/N_c$ corrections, the strict large $N_c$ limit is the same for both the dynamical and quenched theory (with fundamental fermions). More recently, large $N_c$ results from reduced models \cite{Perez:2020fqn,Perez:2020vbn} were made public. We compare the results of the two approaches in Fig. \ref{fig:mesons}. As can be seen, there is a reasonable overall agreement, with differences of at most $2\sigma$. It must be noted that while Ref. \cite{Bali:2013kia} has better statistical accuracy, the results may be affected by significant discretization errors\footnote{A continuum extrapolation is available for $SU(7)$ \cite{Bali:2013fya}.}. In contrast, those of Ref. \cite{Perez:2020vbn} are in the continuum limit.

Another interesting observation is the one of Ref. \cite{Nogradi:2019iek}. The authors found that the ration $M_\rho/F_\pi$ in the chiral limit seemed independent on $N_f$ in $N_c=3$:
\begin{equation}
\frac{M_\rho}{F_\pi} \Bigg \rvert^{N_f=2-6}_{N_c=3} = 7.95(15),
\end{equation}
whereas at large $N_c$, the following result is found \cite{Bali:2013kia}:
\begin{equation}
\sqrt{\frac{N_c}{3}} \frac{M_\rho}{F_\pi} \Bigg \rvert_{N_c\to \infty} = 7.08(10). \label{eq:rhoNc}
\end{equation}
The comparison of the previous equation indicates that both, $1/N^2_c$ and $N_f/N_c$ effects are rather small in this ratio. This is consistent with the result of Ref.~\cite{Nogradi:2019auv}, where the authors summarized the dependence on the gauge group and fermionic representation of the same quantity using all existing lattice data~\cite{Appelquist:2018yqe,Fodor:2016pls,Bali:2013kia,Ayyar:2017qdf,Drach:2017btk,Amato:2018nvj,Bennett:2019cxd,Bennett:2019jzz}. They pointed out that it is almost constant up to a trivial factor that depends on the dimensionality of the fermion fields---$\sqrt{N_c}$ for $SU(N_c)$. We should also mention that the result in eq.~(\ref{eq:rhoNc}) is in the  ballpark of predictions from a resonance chiral theory \cite{Ledwig:2014cla}.

\begin{figure*}[th]
\begin{subfigure}{.5\textwidth}
  \centering
  \includegraphics[width=1\linewidth]{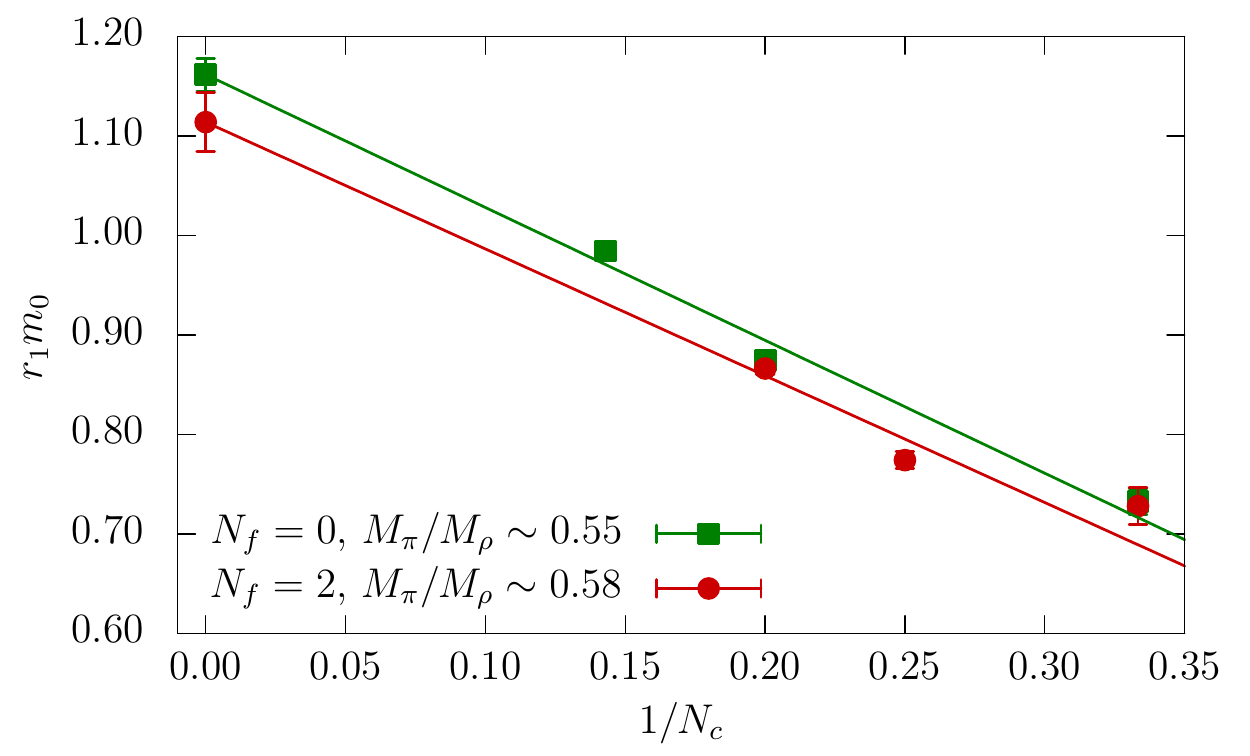}  
  \caption{}
  \label{fig:baryonsm0}
\end{subfigure} 
\begin{subfigure}{.5\textwidth}
  \centering
  \includegraphics[width=1\linewidth]{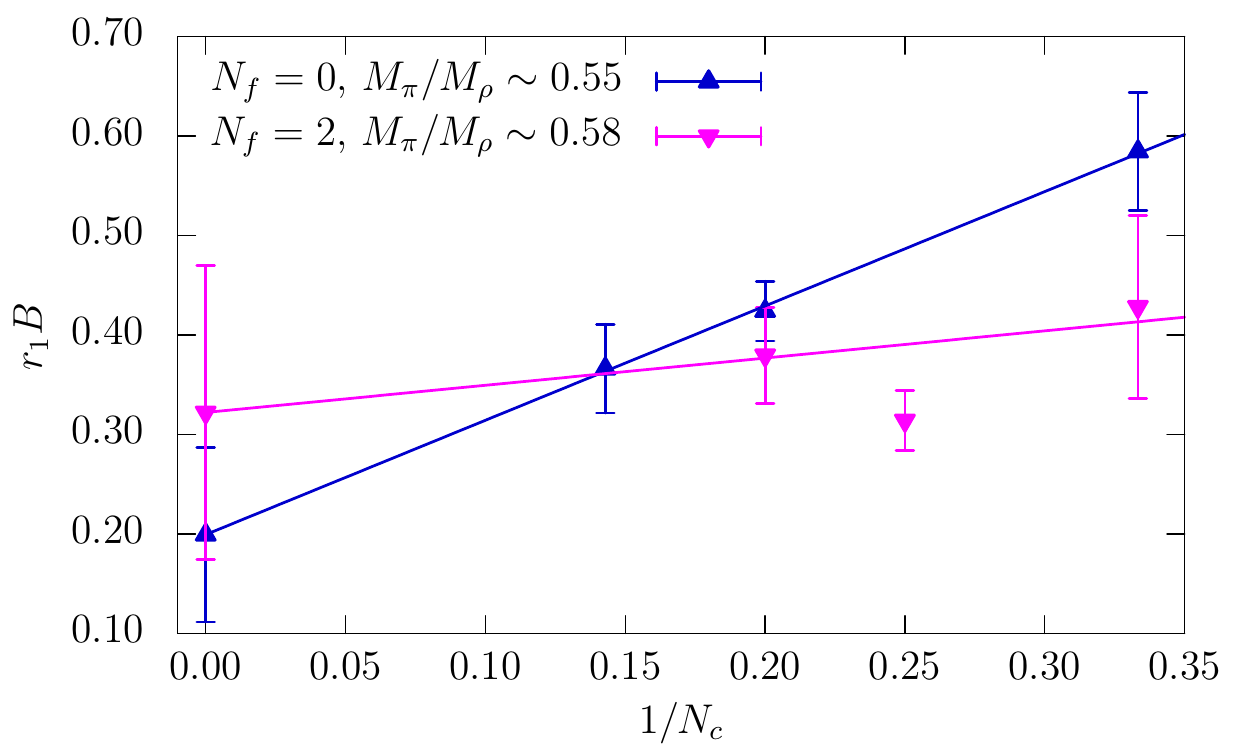}  
  \caption{}
  \label{fig:baryonsB}
\end{subfigure}
\caption{$1/N_c$-scaling of the constants $r_1 m_0$ (top) and $r_1 B$ (bottom) in eq.~(\ref{eq:baryonnf2}) for the quenched data in Ref. \cite{DeGrand:2013nna}, and $N_f=2$ in Ref. \cite{DeGrand:2016pur}. The simulations at different $N_c$ have been matched at approximately fixed $M_\pi / M_\rho$. } 
\label{fig:baryons}
\end{figure*}

\subsection{Baryons}

A series of publications have studied the scaling of baryon masses with the number of colours: first, in the quenched approximation~\cite{DeGrand:2012hd,DeGrand:2013nna,Cordon:2014sda}, and more recently with dynamical quarks ($N_f=2$)~\cite{DeGrand:2016pur}. These have been carried out at a single lattice spacing. In a generic $SU(N_c)$ theory, the spin of the baryon can take half-integer for odd $N_c$ (fermionic baryons, as in QCD), or integer values for even $N_c$ (bosonic baryons).

It has been long known~\cite{Jenkins:1993zu} that baryon masses depend on their angular momentum. This is the so-called hyperfine splitting, which for the case of two degenerate quarks takes the form:
\begin{equation}
M_B(N_c, J) = N_c m_0(N_c) + \frac{J(J+1)}{N_c} B +\cdots, \label{eq:baryonnf2}
\end{equation}
with $m_0, B$ being $O(N_c^0)$ constants. Note that they can have a subleading $N_c$ dependence, for instance, for $m_0$:
\begin{equation}
m_0 = m_0^{(0)} + \frac{m_0^{(1)} }{N_c} + \hdots,
\end{equation}
and similarly for $B$. Generalizations of the eq.~(\ref{eq:baryonnf2}) that incorporate the strange quark also exist in the literature~\cite{Dai:1995zg,Jenkins:1995td}.  The constants in eq.~(\ref{eq:baryonnf2}) can be isolated by taking appropriate combinations, for instance:
\begin{align}
\begin{split}
 m_0 &=  \frac{1}{4 N_c} \left[  5 M(N_c, 1/2) - M(N_c, 3/2)     \right],  \\
B &= \frac{3}{N_c} \left[ M(N_c, 1/2) -M(N_c, 3/2)     \right].
\end{split}
\end{align}

The general conclusions of Refs. \cite{DeGrand:2012hd,DeGrand:2013nna} is that the qualitative expectations of large $N_c$ are satisfied by the data, including those of the $SU(3)$-flavour breaking due to the strange quark. In Ref. \cite{Cordon:2014sda}, combined chiral and $N_c$ fits of the mass and hyperfine splitting were performed, using a consistent expansion in Baryon Chiral Perturbation Theory. That study was able to constrain some of the subleading low-energy constants (LECs).

As an example, we show in Fig. \ref{fig:baryons} the scaling of $m_0$ and $B$ with the number of colours for the quenched and dynamical baryons at approximately fixed Sommer scale $r_1$ \cite{Bernard:2000gd} and $M_\pi/M_\rho$. As can be seen from the plots, quenched and $N_f=2$ results for $r_1 m_0$ agree at fixed $N_c$, and in the large $N_c$ extrapolation. For $r_1 B$, uncertainties are larger, and some quenching effects may be appreciated.  We stress that relative discretization errors may be large, and so, this comparison must be taken as qualitative.

It is also worth mentioning that there have been lattice studies of baryons in the context of BSM physics. These have focused on studying a $SU(4)$ gauge theory with both two fundamental and two sextet fermions~\cite{Appelquist:2014jch,DeGrand:2015lna,Ayyar:2018zuk}. This model represents a slight simplification of the asymptotically-free composite-Higgs Ferretti model \cite{Ferretti:2013kya,Ferretti:2014qta}.

\subsection{Chiral and $N_c$ dependence of light meson observables}

In Ref. \cite{Hernandez:2019qed}, a first-principles calculation of the $N_c$ scaling of the low-energy couplings (LECs) of the chiral Lagrangian has been presented. The LECs are extracted from the chiral dependence of meson masses and decay constants. The lattice setup of this work is $N_f=4$, with $O(a)$-improved Wilson fermions in the sea, and mixed action with twisted mass in the valence. This has the advantage of automatic $O(a)$ improvement, avoiding the need of non-perturbative renormalization factors.

\begin{figure}[ht]
\begin{subfigure}{.5\textwidth}
  \centering
  \includegraphics[width=.999\linewidth]{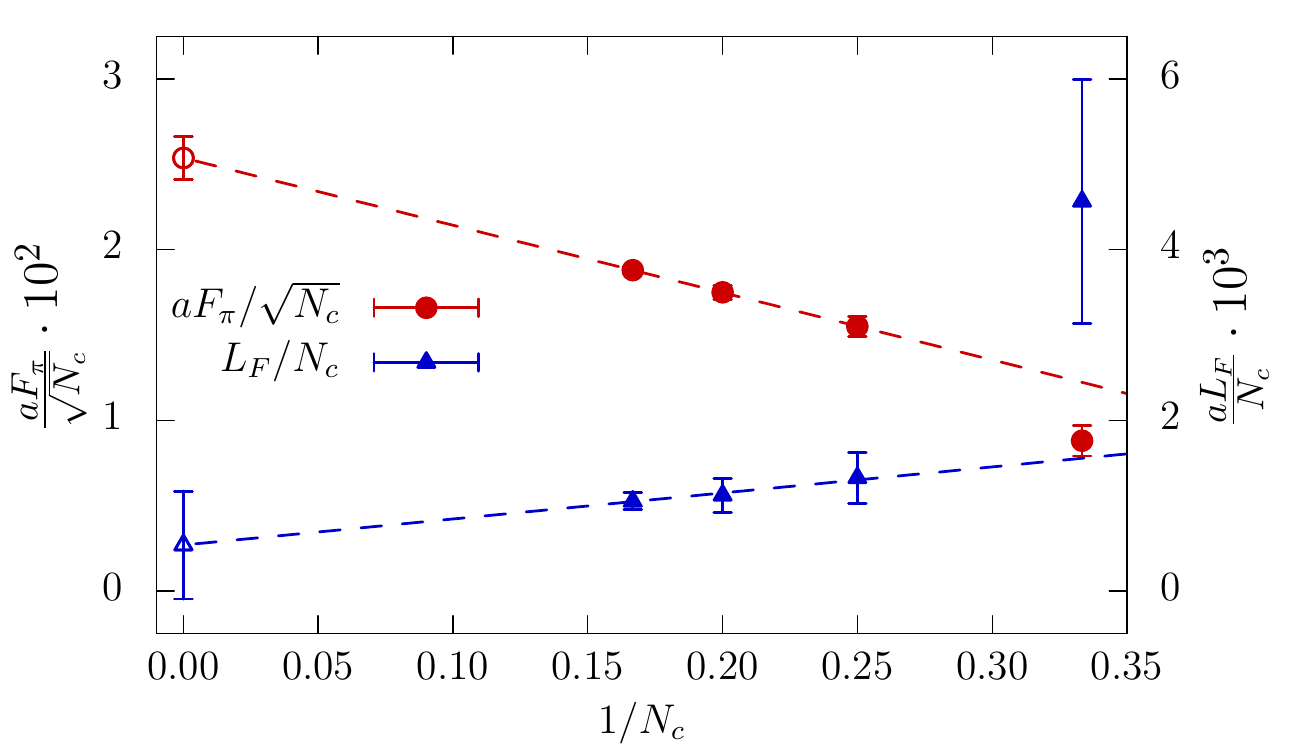}  
  \caption{LECs of the decay constant.}
  \label{fig:fpiscaling}
\end{subfigure} 
\begin{subfigure}{.5\textwidth}
  \centering
  \includegraphics[width=.999\linewidth]{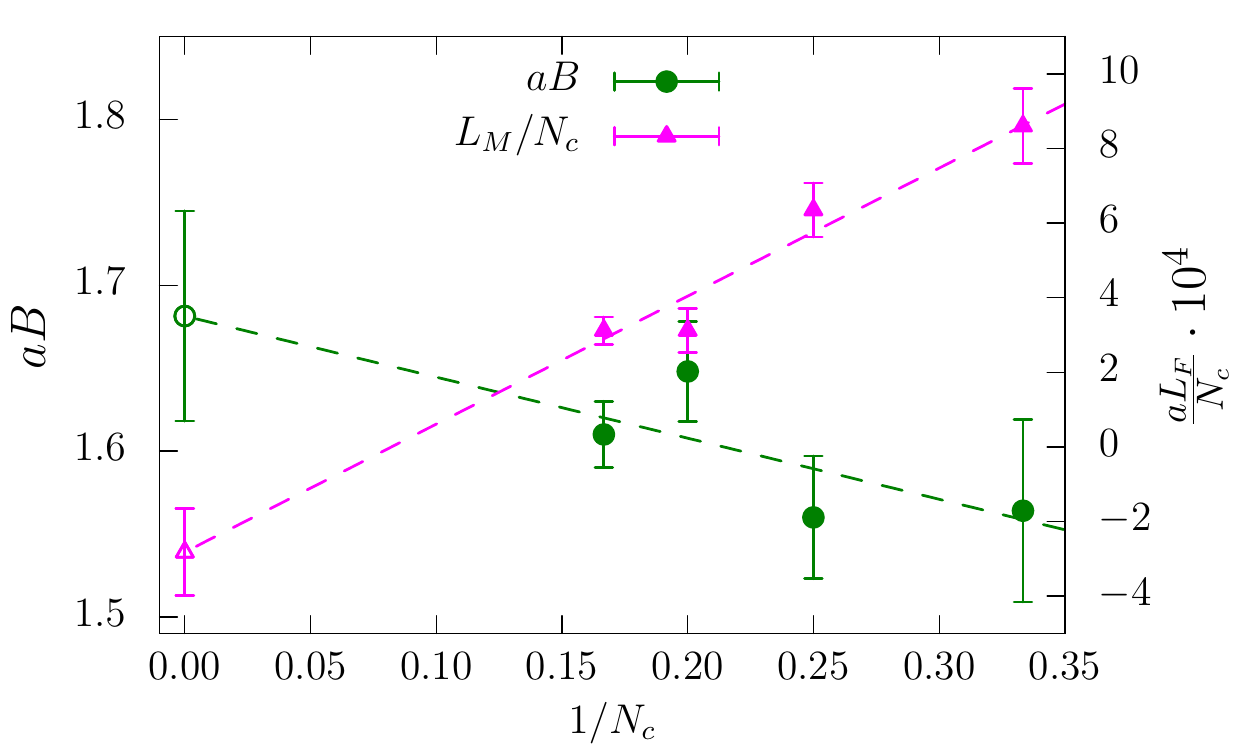}  
  \caption{LECs of the meson mass.}
  \label{fig:massscaling}
\end{subfigure}
\caption{$1/N_c$ scaling of the LO and NLO low-energy constants extracted from fits to eqs. ~(\ref{eq:Fpi1}) and (\ref{eq:Mpi1}) at a fixed value of $N_c$. Source: Ref. \cite{Hernandez:2019qed}.}  
\label{fig:chiral1}
\end{figure}

First the authors consider each value of $N_c$ separately. They perform a chiral fit of the data points to extract the LO and NLO low-energy constants appearing in the NLO chiral predictions \cite{Bijnens1,Bijnens2,Bijnens3} of $F_\pi$ and $M_\pi$: 
\begin{align}
F_\pi &= {F} \Bigg[1   - {\frac{N_f}{2} \frac{M_\pi^2}{(4 \pi F_\pi)^2}\log \frac{M_\pi^2}{\mu^2}}     +4 \frac{M_\pi^2}{F_\pi^2} L_F\Bigg],  
\label{eq:Fpi1}
 \\
\frac{M^2_\pi}{\mu} &= 2B \Bigg[1  + {\frac{1}{N_f} \frac{M_\pi^2}{(4 \pi F_\pi)^2}\log \frac{M_\pi^2}{\mu^2}}  +8\frac{M_\pi^2}{F_\pi^2} L_M\Bigg]. \label{eq:Mpi1} 
\end{align} 
Here, $B$ and $F$ are the chiral condensate\footnote{We note that in the setup of this work, the value of the bare twisted mass can be used for chiral fits as quark mass. The resulting $B$ is therefore also bare.} and decay constant in the chiral limit; $\mu$ is the quark mass, and $L_M$, $L_F$ are combinations the standard LEC:
\begin{equation}
L_F = L_5^r  +  N_f L_4^r , \ \ \ L_M = 2L_8^r - L_5^r + N_f (2 L_6^r  - L_4^r ).
\end{equation}
The scaling with the number of colours of these quantities is known: $F^2, L_5$ and $L_8$ are $O(N_c)$, while $B, L_4$ and $L_6$ are $O(N_c^0)$. The results of these fits are shown in Figs. \ref{fig:fpiscaling} and \ref{fig:massscaling}. One can see that the scaling is well described by leading and subleading $N_c$ corrections for $N_c=4-6$, while there seems to be significant $1/N_c^2$ corrections for $N_c=3$ in the case of  $F$ and $L_F$. In the case of $B$ and $L_M$, there is no sign of $1/N_c^2$ effects.

\begin{figure*}[ht]
\begin{subfigure}{.5\textwidth}
  \centering
  \includegraphics[width=.999\linewidth]{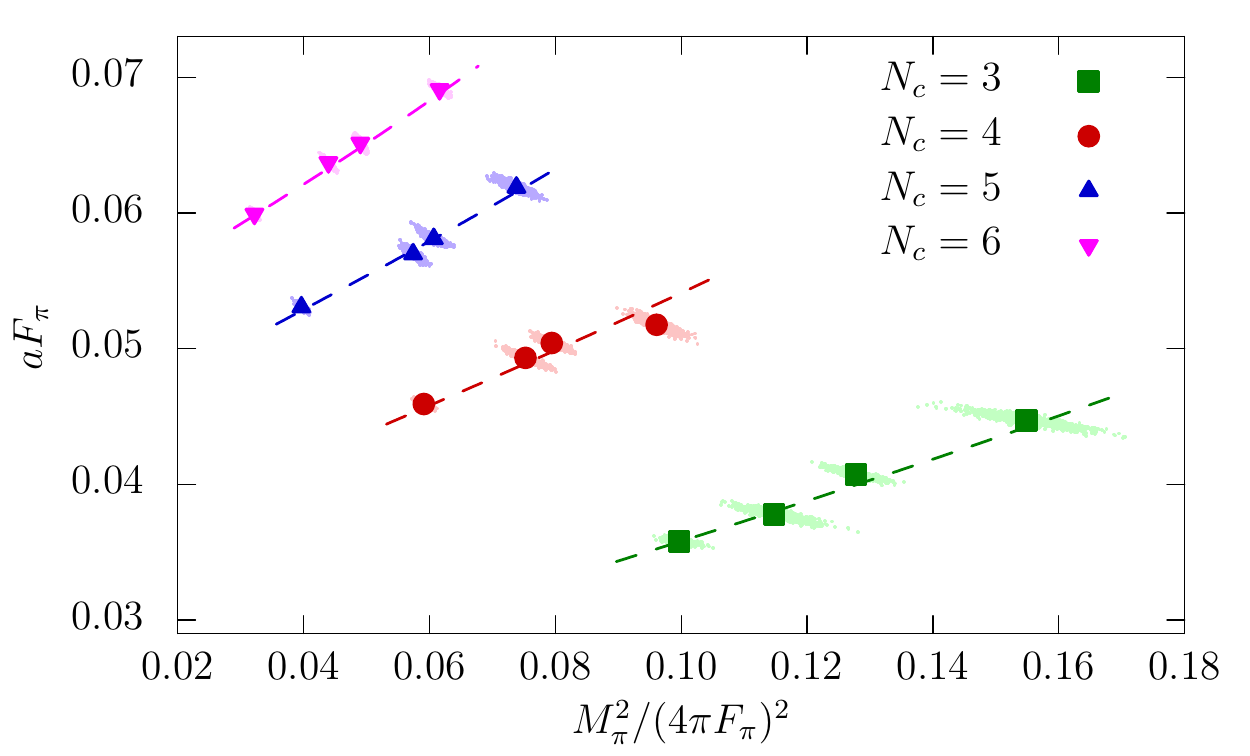}  
  \caption{}
  \label{fig:fpichiral}
\end{subfigure} 
\begin{subfigure}{.5\textwidth}
  \centering
  \includegraphics[width=.999\linewidth]{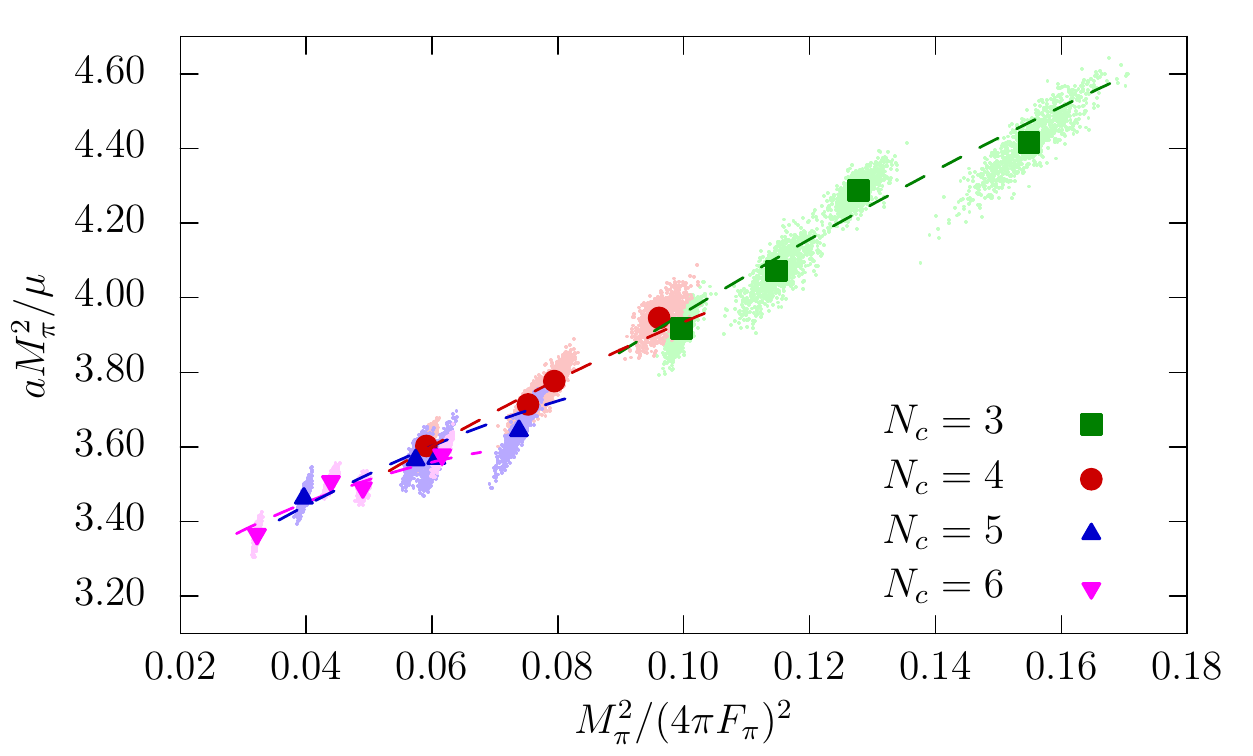}  
  \caption{}
  \label{fig:mpichiral}
\end{subfigure}
\caption{Simultaneous chiral and $N_c$ fits of the decay constant (left) and the pion mass (right). Bootstrap samples are shown as ``clouds'' around the central value. The fit equations are linearized versions of eqs.~(\ref{eq:Mpi}) and (\ref{eq:Fpi}). More detail on the fitting strategy is give in the original article, Ref. \cite{Hernandez:2019qed}.  } 
\label{fig:chiral2}
\end{figure*}

As explained in sec.~\ref{subsec:chpt}, the chiral regime at large $N_c$ requires the addition of the $\eta'$, and a combined power counting of the usual momentum/mass expansion with that in $1/N_c$. Using the standard power-counting \cite{DiVecchia:1980yfw,PhysRevD.21.3388,Witten:1980sp,Kawarabayashi:1980dp,HerreraSiklody:1996pm,Kaiser:2000gs}
\begin{equation}
\mathcal{O}(\delta) \sim  \mathcal{O}(m_q) \sim  \mathcal{O}(p^2) \sim  \mathcal{O}(M_\pi^2) \sim  \mathcal{O}(1/N_c),
\end{equation}
the $\mathcal O (\delta^2)$ predictions  are\footnote{A technical detail in chiral fits is the choice of the renormalization scale. In Ref. \cite{Hernandez:2019qed}, the choice is $\mu^2 = \frac{3}{N_c} (4 \pi F_\pi)^2 $.}:
\begin{eqnarray}
&  F_\pi &=\sqrt{N_c} \left( {F}_0+ {{F}_1\over N_c} + {{ F}_2 \over N_c^2} \right) \Bigg[1    - {\frac{N_f}{2} \frac{M_\pi^2}{(4 \pi F_\pi)^2}\log \frac{M_\pi^2}{\mu^2}} \nonumber \\
&+&4 \frac{M_\pi^2}{F_\pi^2} \Big( N_c {L}_F^{(0)} + {L}_F^{(1)} \Big) + N_c^2 K_F^{(0)} \left({M_\pi^2\over F_\pi^2}\right)^2 \\&+& \ {\mathcal O}(\delta^3)  \Bigg] , \nonumber
\label{eq:Fpi}
\end{eqnarray}
and 
\begin{eqnarray}
&M^2_\pi &= 2  m \left( { B}_0+ {{ B}_1\over N_c} + {{B}_2 \over N_c^2} \right)  \Bigg[ \nonumber  \\
&1&+ {\frac{1}{N_f} \frac{M_\pi^2}{(4 \pi F_\pi)^2}\log \frac{M_\pi^2}{\mu^2}} - {\frac{1}{N_f} \frac{M_{\eta'}^2}{(4 \pi F_\pi)^2}\log \frac{M_{\eta'}^2}{\mu^2}}\\ & +&8\frac{M_\pi^2}{F_\pi^2} \Big( N_c {L}_M^{(0)} + {L}_M^{(1)} \Big) + N_c^2 K_M^{(0)} \left({M_\pi^2\over F_\pi^2}\right)^2 +{\mathcal O}(\delta^3) \Bigg] \nonumber  .\label{eq:Mpi}
\end{eqnarray}
Here, $F_i, B_i, L_M^{(i)}$ and $L_F^{(i)}$ are the coefficients of the $1/N_c$ expansion of the corresponding LECs. $K_{F,M}$ are combinations of LECs that contribute at  $\mathcal{O}(M_\pi^4)$.\footnote{For further details see \cite{Guo:2015xva}. }. A global chiral fit of all $N_c$ data to these expressions is shown in Fig. \ref{fig:chiral2}. The ChPT predictions seem to describe data well. 

An interesting observation of Ref. \cite{Hernandez:2019qed} is that by studying the first coefficients in the $1/N_c$ expansion, one can infer the values of some quantities at different value of $N_f$, since the leading corrections come from fermion loops and are therefore of the form $N_f/N_c$ (the non-planar gluonic corrections start at $1/N_c^2$). The simplest example is the decay constant in the chiral limit, whose leading $N_f,N_c$ dependence is:
\begin{equation}
\frac{F}{\sqrt{N_c}} = F_0 +  \widetilde F_1 \frac{N_f}{N_c} + \cdots.
\end{equation}
Since $\widetilde F_1$ is trivially related by a factor $N_f$ to one of the fit parameters in eq.~(\ref{eq:Fpi}), the authors are able to quote inferred results for $N_f=2$ and $3$:
\begin{align}
\begin{split}
& F^{N_c=3, N_f=2} = 81(7)  \text{ MeV},  \\
& F^{N_c=3, N_f=3} = 68(7)  \text{ MeV},\label{eq:FNf2}
\end{split}
\end{align}
which are in good agreement with phenomenological and lattice determinations---see Ref. \cite{Aoki:2019cca}. The same strategy is followed for the LECs, and the chiral condensate.

We should also mention that the chiral behaviour of $F_\pi$ in $N_f=2$ simulations (and the equivalent for vector meson) was also studied at in Ref. \cite{DeGrand:2016pur}, albeit no chiral fits were performed.

Another quantity that may be explored are meson scattering amplitudes. Although they vanish in the exact large $N_c$ limit, it is interesting to study the subleading $N_c$ behaviour, and the interplay with the quark mass dependence. For instance, the isospin-2 leading order ChPT prediction is\footnote{For the scattering length, we are using the convention $k \cot \delta_0 = -1/a_0 + O(k^2)$.}
\begin{equation}
M_\pi a_0^{I=2} = \frac{M_\pi^2}{16 \pi F_\pi^2} \sim O\left(\frac{1}{N_c} \right), \label{eq:chpta0}
\end{equation} 
that is consistent with the $N_c$-scaling discussed in sec.~\ref{subsec:chpt}.

 In finite volume, scattering amplitudes may be obtained with the Lüscher formalism~\cite{Luscher:1986pf}, and generalizations thereof (see Ref. \cite{Briceno:2017max} for a review). In the so-called threshold (or $1/L$) expansion, valid for weakly interacting systems, it is possible to relate the finite-volume energy shift of the two-particle ground state to the scattering length by to the scattering length, see Ref. \cite{Luscher:1986pf}.

Recently,  preliminary results   for the isospin-2 scattering length at $N_c=3-6$ have been presented~\cite{Romero-Lopez:2019gqt}. As $N_c$ grows, two different effects compete. The signal for the energy shift becomes weaker as the scattering length decreases, but the statistical error gets reduced as a result of the increase of gluonic degrees of freedom. In Ref. ~\cite{Romero-Lopez:2019gqt}  $a_0^{I=2}$ at $N_c>3$ was obtained with good accuracy, and a remarkable agreement with leading order ChPT [see eq. (\ref{eq:chpta0})] was found. 

Additional channels, such as the isospin-0 ($\sigma$) channel may represent an interesting problem to study in the large $N_c$ context. It has been argued that the $N_c$-scaling may shed light on the nature of the $\sigma$ resonance~\cite{Pelaez:2010er,Nebreda:2011cp,Bernard:2010fp,RuizdeElvira:2017aet}. Another very interesting application would be the study of tetraquarks with different flavour content in the planar limit \cite{PhysRevLett.110.261601,PhysRevD.88.036016}. 

Finally, we should mention the results for the $\rho$\footnote{The $\rho$ resonance has been treated as a stable state in Refs. \cite{Bali:2013kia,Nogradi:2019iek,DeGrand:2016pur}. This is rigorous for heavier pion masses, but implies systematic errors for lighter pions.}  and the isospin-2 channels in $SU(2)$\cite{Janowski:2019svg,Arthur:2014zda}. Note that in this case the symmetry breaking pattern for $N_c=2$ is different than the one in $N_c>3$, and so effective field theory expectations differ.

\subsection{Topological susceptibility with dynamical quarks }

The topological susceptibility is an observable that has a very different behaviour in the pure Yang-Mills theory and in QCD with light quarks. In the former, discussed in sec. \ref{subsec:topsus1}, $\chi_{YM}$ depends only weakly in the number of colours; whereas in the latter, it vanishes in the chiral limit for all values of $N_c$.  It has been predicted \cite{DiVecchia:1980yfw,Leutwyler:1992yt,Crewther:1977ce} that the topological susceptibility mass-dependence is as follows:
\begin{equation}
\frac{1}{\chi(M_\pi)} = \frac{2N_f}{F^2_\pi M_\pi^2 } + \frac{1}{\chi_{YM}}. \label{eq:topsusdyn}
\end{equation}
In the previous equation, the pure gauge (quenched) limit is achieved by letting $M_\pi \to \infty$, whereas at small pion mass, one simply has $\chi(M_\pi) \propto M_\pi^2$.

\begin{figure}[h!]
\centering 
\includegraphics[width=0.5\textwidth]{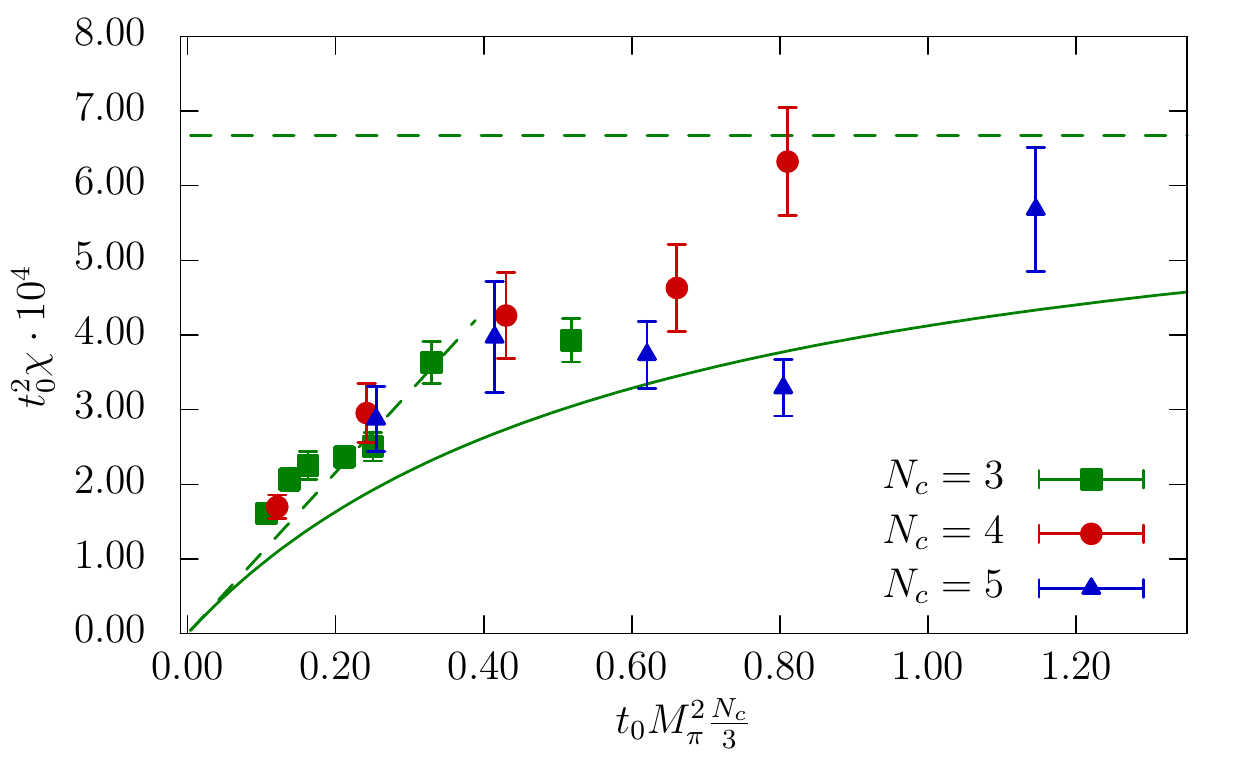}
\caption{Topological susceptibility as a function of the pion mass in $N_f=2$ QCD. The solid line is the functional form in eq.~(\ref{eq:topsusdyn}), using the continuum results of Refs. \cite{Ce:2015qha,Bruno:2014ova}. The dashed lines are the expectations in the two regimes (quenched and light pions). The various coloured points are the results of Ref. \cite{DeGrand:2020utq} with $N_c=3-5$. \label{fig:topsusdyn}}
\end{figure}

The topological susceptibility with dynamical fermions has been studied for $N_c=3$ \cite{Durr:2001ty,Chiu:2011dz,Bruno:2014ova,Aoki:2007pw}.  The most recent result \cite{Bruno:2014ova}, includes a continuum limit extrapolation. In the present year, a ``pilot study'' with varying $N_c=3-5$~\cite{DeGrand:2020utq} was also published.

Combining the results in the Yang-Mills theory \cite{Ce:2015qha}, with the $N_f=2$ ones \cite{Bruno:2014ova} one can produce an interpolating curve between the two regimes based on eq.~(\ref{eq:topsusdyn}). This can be compared to the results of Ref.~\cite{DeGrand:2020utq} with $N_c=3-5$. As can be seen, qualitative (but not quantitative) agreement is observed. One possibility to explain the disagreement are discrezation effects, which were found in Ref. \cite{Bruno:2014ova}  to be relevant with the Wilson fermionic action. Finite volume effects may also be significant in Ref.~\cite{DeGrand:2020utq}.

\subsection{The $\Delta I =1/2$ rule}

In the last few years, impressive progress has been achieved in the lattice determination of the $K\rightarrow \pi\pi$ amplitudes, and the related  CP violation observable $\epsilon'/\epsilon$~\cite{Boyle:2012ys,Bai:2015nea,Blum:2015ywa,Ishizuka:2018qbn,Abbott:2020hxn}.  The $\Delta I=1/2$ enhancement seems to be reproduced by  the latest simulations at the physical point in $2+1$ simulations with a heavy charm \cite{Abbott:2020hxn}. 

In an earlier work \cite{Boyle:2012ys}, an analysis of the various contributions to $K \to \pi\pi$   suggested that the main source of the enhancement comes from the current-current operators, eqs.~(\ref{eq:currento}). It emerges from a strong cancellation of the isospin-2 amplitude, as a result of a negative relative sign between the colour-connected and colour-disconnected contractions, which scale differently in large $N_c$. The scaling in $N_c$ of these amplitudes allows therefore to disentangle the two contributions rigorously, and has been studied in Refs. \cite{Donini:2016lwz,Donini:2020qfu}.

In these studies  an old strategy~\cite{Giusti:2004an,Giusti:2006mh} was revisited in the context of large $N_c$, where a light active charm, mass-degenerate with the up quark, was included, simplifying enormously the calculation. In this scenario, only the two current-current operators, $Q^\pm$, fully describe the $\Delta S = 1$ transitions. A matching to $N_f=4$ chiral perturbation theory, allows to extract the low-energy couplings, $g^\pm$, representing the strength of these interactions in the chiral Lagrangian, eq.~(\ref{eq:chpt4}). The couplings can be extracted from the measurement of the simpler $K \to \pi$ amplitudes, $A^\pm$ in the chiral limit. The determination of $g^\pm$ gives then an indirect estimate of the $K \to \pi \pi$ isospin amplitudes.

\begin{figure}[h!]
\begin{subfigure}{.5\textwidth}
  \centering
  \includegraphics[width=.999\linewidth]{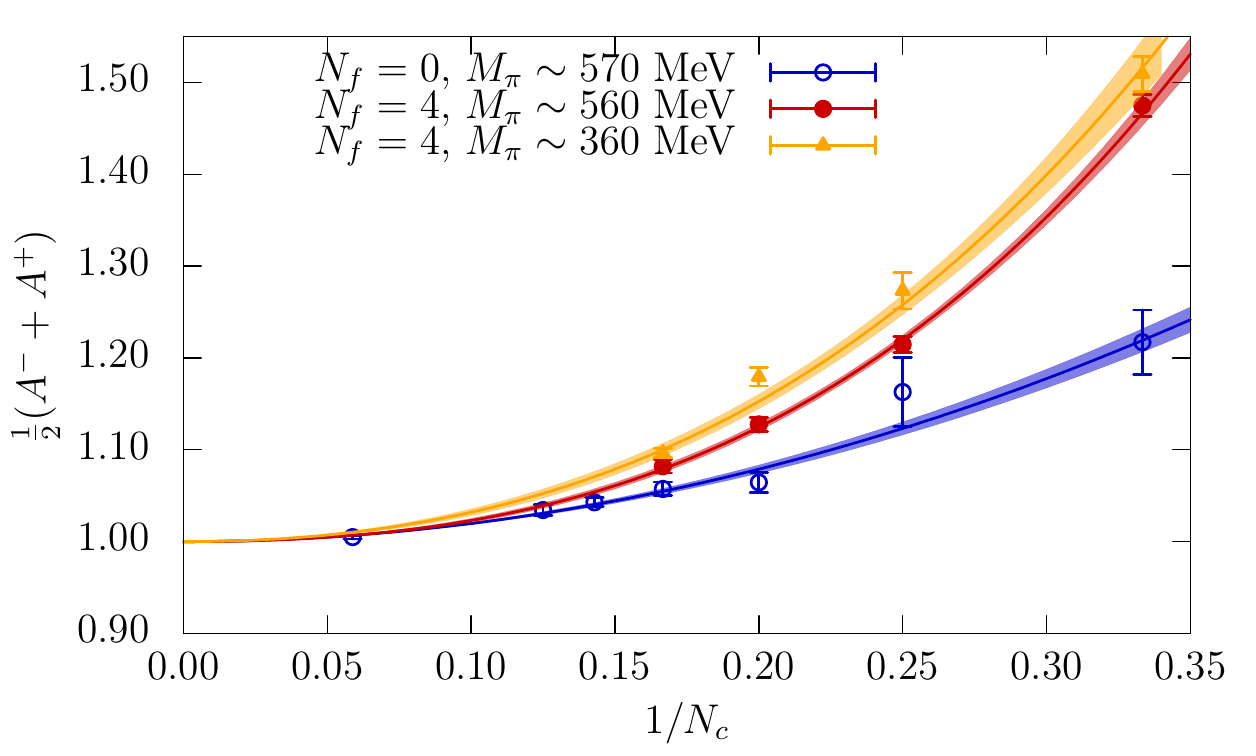}  
  \caption{}
  \label{fig:ammap}
\end{subfigure}
\begin{subfigure}{.5\textwidth}
  \centering
  \includegraphics[width=.999\linewidth]{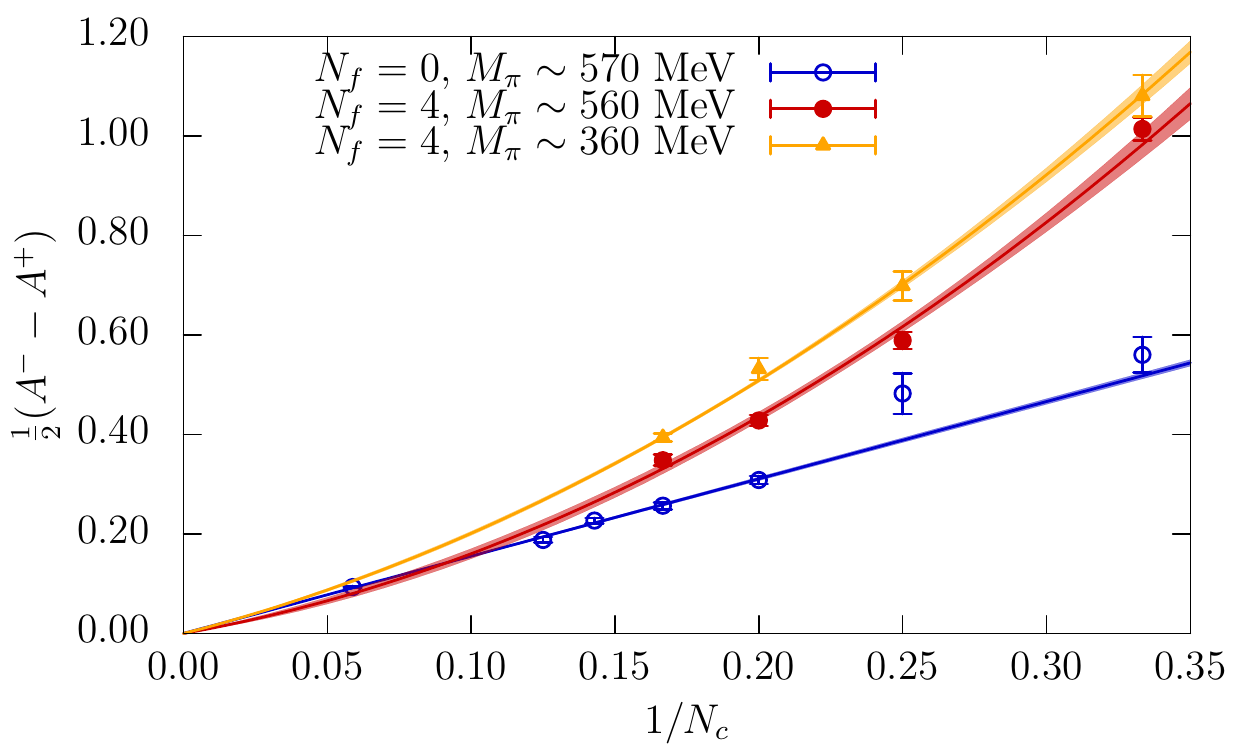}  
  \caption{}
  \label{fig:ampap}
\end{subfigure}
\caption{Half-sum and half-difference of the amplitudes $A^\pm$ as a function of $N_c$ for three different cases: (i) quenched results in blue, (ii) dynamical results at a pion similar to the quenched case (red), and (iii) dynamical results at a lighter pion mass (orange). Error bars include only statistical errors. Source: Refs. \cite{Donini:2016lwz,Donini:2020qfu}. } 
\label{fig:ktopi1}
\end{figure}

\begin{figure*}[ht]
\begin{subfigure}{.5\textwidth}
  \centering
  \includegraphics[width=.999\linewidth]{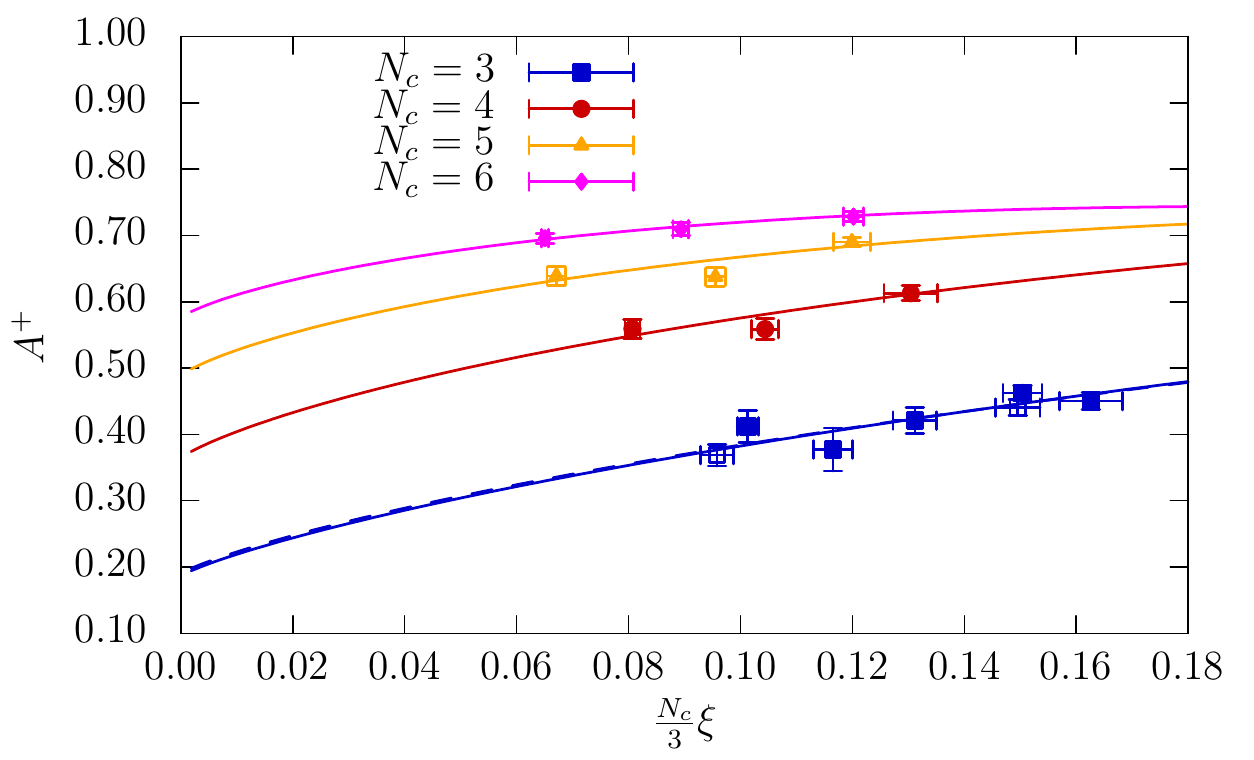}  
  \caption{}
  \label{fig:plotplus}
\end{subfigure}
\begin{subfigure}{.5\textwidth}
  \centering
  \includegraphics[width=.999\linewidth]{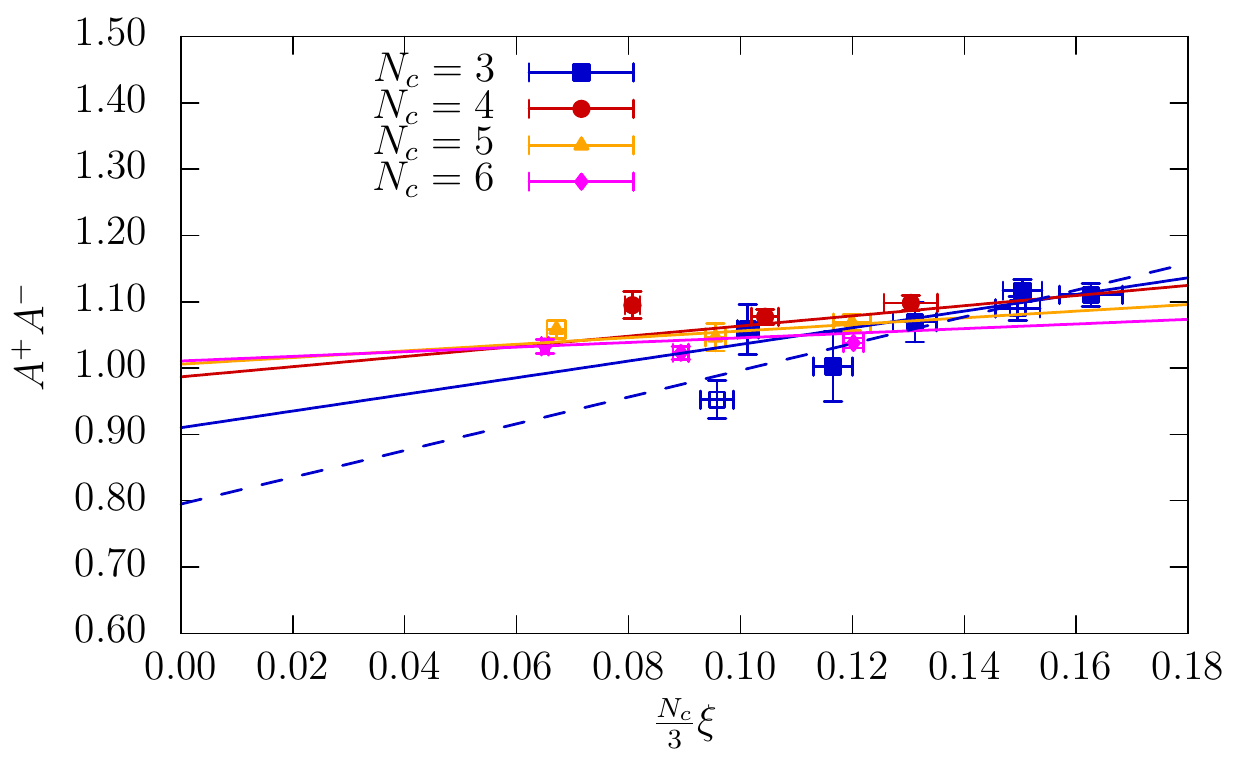}  
  \caption{}
  \label{fig:plotprod}
\end{subfigure}
\caption{Chiral extrapolation of $A^+$ and the product $A^+A^-$. Empty squares for $N_c=3$ indicate a finer lattice spacing. Solid lines indicate a simultaneous chiral and $N_c$ fit. Dashed lines represent the chiral extrapolation of the data points with $N_c = 3$. Source: Ref. \cite{Donini:2020qfu},} 
\label{fig:ktopi2}
\end{figure*}

The expected large $N_c$ scaling of the amplitudes $A^\pm$ has been presented in eq.~(\ref{eq:rnc}).
By means of appropriate linear combinations, one can isolate the (anti-)correlated coefficients:
\begin{align}
\frac{A^- + A^+}{2} &= 1 + \tilde c \frac{1}{N_c^2}  + \tilde d \frac{N_f}{N_c^3} , \\ \frac{A^- - A^+}{2} &=  \tilde a \frac{1}{N_c}+\tilde b \frac{N_f}{N_c^2}.
\end{align}
The $N_c$ scaling of these combinations was analysed in Ref. \cite{Donini:2020qfu} in three different scenarios: (i) quenched results ($N_f = 0$) at a heavy pion mass $\sim 570$ MeV, (ii) dynamical results ($N_f = 4$) at a heavy pion mass $\sim 560$ MeV, and (iii) dynamical results ($N_f = 4$) at a lighter pion mass $\sim 360$ MeV. This can be seen in Fig. \ref{fig:ktopi1} and Table \ref{tab:fitsNc}. The main conclusion from this study is that all coefficients turned out to be of the natural size. Importantly, the sign of the $\tilde a$ and  $\tilde b$ coefficients is the same and negative. This implies that both terms contribute to reduce the $A^+$ amplitude and enhance, in a correlated way, the amplitude $A^-$. Moreover, it can be seen that the mass dependence for the $N_f = 4$ results seems to affect mostly the coefficient $\tilde a$, and goes also in the direction of enhancing the ratio $A^-/A^+$ towards the chiral limit.

\begin{table}[h!]
\centering
\begin{tabular}{c|c|c|c|c}
\multicolumn{5}{c}{ Half-difference} \\ \hline
    \rule{0pt}{10.5pt}    Case            & $M_\pi$      & $\tilde{a}$ & $\tilde{b}$ & $\chi^2/\text{d.o.f.}$ \\ \hline\hline
$N_f=0$& $570$ MeV  &  -1.55(2)           &      ---       &8.8/6   \\ \hline
$N_f=4$& $560$  MeV &  -1.03(13)           &   -1.44(13)       & 6.6/2     \\ \hline
$N_f=4$& $360$ MeV &     -1.49(15)        &   -1.32(18)      & 0.3/2      \\ \hline
\end{tabular}

\vspace{0.5cm}

\begin{tabular}{c|c|c|c|c}
\multicolumn{5}{c}{ Half-sum} \\  \hline
  \rule{0pt}{10.5pt}    Case           & $M_\pi$      &$\tilde{c}$ & $\tilde{d}$& $\chi^2/\text{d.o.f.}$ \\ \hline\hline 
$N_f=0$& $570$   MeV&     2.1(1)       &     ---      & 3.5/6  \\ \hline
$N_f=4$& $560$   MeV&\  \ 1.2(3) \  \       &\  \ 2.2(3) \ \ &    1.3/2     \\ \hline
$N_f=4$& $360$  MeV& \  \  2.4(4) \   \    &\  \  1.6(4) \ \ &  3.2/2      \\ \hline
\end{tabular}

\caption{Summary of results for the $1/N_c$ fits to the half-sum and half-difference of the amplitudes $A^\pm$. Errors are only statistical.}
\label{tab:fitsNc}
\end{table}

The determination of the couplings $g^\pm$ requires an extrapolation to the chiral limit. The ChPT prediction for these amplitudes is \cite{Hernandez:2006kz,Kambor:1989tz}:
\begin{equation}
A^\pm = g^\pm \left[ 1 + 3 \left( \frac{M_\pi}{4 \pi F_\pi}\right)^2\log \frac{M_\pi^2}{\Lambda_\pm^2} \right]. 
\end{equation}
The chiral fits are shown in Fig. \ref{fig:ktopi2} for the $A^+$ amplitude (Fig. \ref{fig:plotplus}) and the product of the two $A^+ A^-$ (Fig. \ref{fig:plotprod}). The result for $N_c=3$ is:
\begin{equation}
\frac{g^-}{g^+} = 22(5),
\end{equation}
where the error is only statistical. Finally, an indirect estimate of the $K \to \pi \pi$ isospin amplitudes can be derived using the LO ChPT prediction in eq.~(\ref{eq:A0overA2}), and also the NLO
one derived in Ref. \cite{Donini:2020qfu}. The final estimate of Ref. \cite{Donini:2020qfu} for this ratio in the theory with a light charm is
\begin{equation}
\frac{A_0}{A_2} \Bigg \rvert_{N_f=4, N_c=3} = 24(5)_\text{stat}(7)_\text{sys}. \label{eq:A0A2}
\end{equation}

The main conclusion from this work is that the large enhancement observed in the $\Delta I=1/2$ rule seems consistent with coefficients in the $1/N_c$ expansion that are of the natural size, but with an important effect coming from the quark loops.  In addition, the result in eq.~(\ref{eq:A0A2}) suggests that the enhancement may indeed be largely dominated by intrinsic QCD effects, and not from the intermediate scale set by the charm quark mass.

\section{Concluding remarks}
We have reviewed recent lattice studies of the large $N_c$ limit of QCD. Our main focus has been the study of large $N_c$ scaling of various physical quantities at zero temperature, using the conventional approach of comparing standard lattice simulations at physical volumes and increasing values of $N_c$. Some of the results obtained with non-standard approaches, such as the Eguchi-Kawai reduction (and variations thereof),  are also included in the comparison of results.

We have presented results in pure Yang-Mills, more specifically, a non-perturbative study of factorization, the glueball spectrum and the topological susceptibility. In the case of theories with fermions, results on the meson and baryon spectrum, the chiral behaviour of meson masses and decay constants, the topological susceptibility and some preliminary results on $\pi\pi$ scattering have been reviewed.  We discussed in some detail a compelling new result on the $1/N_c$ scaling of weak amplitudes related to the $\Delta I=1/2$ rule in kaon decays. 

In most of these cases, the expected scaling in $N_c$ is confirmed.  Coefficients of the natural size are found, that is ${\mathcal O}(N_c^{-1})$ relative corrections. In particular, this is found to be the case  for the ratio of isospin amplitudes in kaon decays. In some cases, however, the coefficients of the subleading corrections might be significantly enhanced when leading and subleading corrections do measure different physics scales.  This is for example the case when studying violations to factorization.
 
 In this respect one interesting direction for future work is the study of meson interactions with growing $N_c$. This involves not only pion scattering, but also glueball interactions and potential exotic resonances, such as tetraquarks, that may prevail at large $N_c$. 

Finally we should mention other works that we have not reviewed, for instance, Ref. \cite{Athenodorou:2017cmw}. A number of results at varying $N_c$ at finite-temperature exist\footnote{See Ref. \cite{Lucini:2012gg} for a discussion of earlier results on finite-temperature at large $N_c$.} \cite{Lucini:2003zr,DelDebbio:2004vxo,Lucini:2005vg,Lucini:2012wq,DeGrand:2018tzn,Bonati:2013tt,Datta:2010sq,Bringoltz:2005rr,Bonati:2019unv,Bonati:2019kmf,Bursa:2005yv,Panero:2009tv,Mykkanen:2012ri}. Recent lattice results with different gauge groups and fermion content can be found in \cite{Holligan:2019lma,Bennett:2020hqd,Ayyar:2018glg,Brower:2020mab}.

\vspace*{1.5cm}
\noindent
\centerline{\textbf{Acknowledgements}}
\vspace*{0.1cm}

We warmly thank T. DeGrand, M. Garc{\'i}a-P{\'e}rez, M. Garc{\'i}a Vera, C. Pena, A. Ramos and R. Sommer for useful discussions. 

We acknowledge support from the Generalitat Valenciana grant PROMETEO/2019/083, the European project H2020-MSCA-ITN-2019//860881-HIDDeN, and the Spanish project FPA2017-85985-P.

The work of FRL has also received funding from the EU Horizon 2020 research and innovation program under the Marie Sk{\l}odowska-Curie grant agreement No. 713673 and La Caixa Foundation (ID 100010434). FRL also acknowledges financial support from Generalitat Valenciana through the plan GenT program (CIDEGENT/2019/040).

\vspace*{.5cm}
 \bibliographystyle{epj}
 \bibliography{references}

\end{document}